\documentclass[12pt]{article}
\usepackage{amsmath}
\usepackage{amssymb}
\usepackage{graphicx}
\usepackage{enumerate}
\usepackage{natbib}
\usepackage{url} 
\usepackage{bbm}
\usepackage{comment}
\usepackage{commath}
\usepackage{blkarray}
\usepackage{xcolor}
\usepackage{booktabs}
\usepackage{geometry}

\newcommand{\blind}{1}

\begin{document}

\def\spacingset#1{\renewcommand{\baselinestretch}%
{#1}\small\normalsize} \spacingset{1}


\if1\blind
{
  \title{\bf Statistical modeling of on-street parking lot occupancy in smart cities}
\author{Marc Schneble\\
    Department of Statistics, Ludwig-Maximilians-Universit\"{a}t Munich\\
    and \\
    G\"{o}ran Kauermann \\
    Department of Statistics, Ludwig-Maximilians-Universit\"{a}t Munich}
  \maketitle
} \fi

\if0\blind
{
  \bigskip
  \bigskip
  \bigskip
  \begin{center}
    {\LARGE\bf Statistical modeling of on-street parking lot occupancy in smart cities}
\end{center}
  \medskip
} \fi

\bigskip
\begin{abstract}
Many studies suggest that searching for parking is associated with significant direct and indirect costs. Therefore, it is appealing to reduce the time which car drivers spend on finding an available parking lot, especially in urban areas where the space for all road users is limited. The prediction of on-street parking lot occupancy can provide drivers a guidance where clear parking lots are likely to be found. This field of research has gained more and more attention in the last decade through the increasing availability of real-time parking lot occupancy data. In this paper, we pursue a statistical approach for the prediction of parking lot occupancy, where we make use of time to event models and semi-Markov process theory. The latter involves the employment of Laplace transformations as well as their inversion which is an ambitious numerical task. We apply our methodology to data from the City of Melbourne in Australia. Our main result is that the semi-Markov model outperforms a Markov model in terms of both true negative rate and true positive rate while this is essentially achieved by respecting the current duration which a parking lot already sojourns in its initial state.
\end{abstract}

\noindent%
{\it Keywords:}  (Inverse) Laplace transform; predicting parking lot occupancy; semi-Markov process; time to event analysis
\vfill

\newpage
\spacingset{1} 

	\section{Introduction}

Finding a clear parking lot in the center of a metropolitan region is usually very time-consuming and hence an expensive affair. A recent study that estimates the ``economic and non-economic impact of parking pain'' is \cite{cooksoninrix}. The study is mainly concerned with 10 major cities in each, the United States, the United Kingdom and Germany, respectively, and states that the average annual search time for parking ranges from 35 to 107 hours. This induces direct and indirect costs of up to 2,243 US-Dollars per year on the individual basis. The economic costs comprise the costs for fuel and opportunity costs of wasted time whereupon the non-economic costs are, among other things, related to higher stress levels. Other studies measured the share of traffic cruising for parking  and quantified the average duration until a parking lot is found (e.g. \citeauthor{shoup2017high}, \citeyear{shoup2017high} or \citeauthor{cao2017impacts}, \citeyear{cao2017impacts}). In \cite{hampshire2018share}, the authors compare the results of 22 of these studies. The share of traffic which cruises in order to find parking ranges from 8\% to 74\%, where the percentages depend heavily on the location and the time of the day. However, most of the studies suffer from a selection bias as they are oftentimes focused on regions where the demand for parking is generally very high. In any case, the search for parking enhances traffic congestion which itself causes an increasing number of accidents, air pollution, noise, etc. (\citeauthor{goodwin2004economic}, \citeyear{goodwin2004economic}).

Car drivers could reduce all the costs and harms mentioned above if they would know ahead of time where the chance of finding a free parking lot is greatest. By the use of wireless sensor technologies (e.g. \citeauthor{lee2008intelligent}, \citeyear{lee2008intelligent}), so called ``smart cities'' (\citeauthor{lin2017survey}, \citeyear{lin2017survey}) are able to collect information regarding parking lot occupancy in real time. This information can be supplied to the public via smartphone apps or a direct gateway to the car. In general, one can either measure the number of free parking lots currently available in a predefined area, e.g. in the parking garage of a mall, or one can measure the occupancy of each single parking lot, e.g. for on-street parking. In this paper, we focus on the latter. A non-exclusive list of cities which already have implemented public accessible on-street parking sensors comprises San Francisco, California \citep{saharan2020efficient}, Santander, Spain \citep{cheng2015building} and Melbourne, Australia \citep{melbourne2020}. A summary of smart parking city projects around the world is provided by \cite{lin2017survey}.   

For both off-street and on-street parking, various methodologies have been developed which aim at the prediction of parking lot occupancy by leveraging the data collected by smart cities. Real time parking data from San Francisco are employed by \cite{rajabioun2015street} in order to predict the spatio-temporal pattern of parking availability via a multivariate autoregressive model. Neural networks for parking availability prediction are used by \cite{zheng2015parking} and \cite{vlahogianni2016real}. The former paper defines a set of previous observations, the calendar time and the day of the week as input variables. Regression trees and support vector regression are used as comparative methods. The latter paper makes use of a simple neural network model for time series prediction with a specified number of lagged parking occupancy rates and with application to data from the city of Santander.  An extension of a parking prediction model to an online parking guidance system was designed by \cite{liu2018street}. Their model can also handle multiple users looking simultaneously for a free parking lot. The availability of parking is modeled via an autoregressive model and the recommended parking lot is a linear function of both driving cost and walking cost.  Deep learning with recurrent neural networks are utilized by \cite{camero2018evolutionary} to predict occupancy rates of car parks in Birmingham. They compare their results in terms of mean absolute error with already existing prediction techniques on the same data set. Among others, a time series approach lead to higher prediction accuracy. 

A more statistical approach to predict off-street parking occupancy  was outlined first in \cite{caliskan2007predicting} and revisited by \cite{klappenecker2014finding}. Both papers model each car park as a queue which is described by a continuous time Markov process, i.e. the duration times in each state are assumed to be exponentially distributed. In particular, the transition matrix is dependent on two parameters, the arrival rate and the parking rate which are both assumed to be constant over time. Moreover, in both papers the model is evaluated only with simulated data which does not answer the question of whether the model is actually able to reliably predict parking lot availability. In a similar manner, \cite{monteiro2018street} propose to model the arrival and the departure rate at on-street parking lots via non-homogeneous Poisson processes.

In this paper, we follow up the idea of modeling parking lot occupancy as a stochastic process. Since we concern ourselves with on-street parking only, we model each parking lot as a two-state stochastic process. As it turns out, semi-Markov processes \citep{pyke1961markov}, which allow non-exponentially distributed duration times, are an appropriate class of stochastic processes for our study. These kinds of processes have wide-ranging applications, e.g. in production systems and maintenance systems where the time spend in an operational state is of interest \citep{limnios2012semi}. In order to estimate the transition intensities of the semi-Markov process we employ time to event models which are essentially used in the epidemiological field (see e.g. \citealp{klein2006survival} and \citealp{kalbfleisch2011statistical}). We thereby respect the spatial dependence of nearby parking lots as well as further unobserved parking lot specific heterogeneity by including random effects in the model.

  
The remainder of this paper is organized as follows. In Section \ref{sec:data} we visualize the data from the City of Melbourne which we use for our analyses and already provide some descriptive statistics. In Section \ref{sec:notation} we introduce some notation and state the problem that we tackle in this paper. Sections \ref{sec:semiMarkov} and \ref{sec:intensities} are concerned with the methodology  involving  semi-Markov processes and time to event analysis.  The results when applying our methodology to the Melbourne parking data are presented in Section \ref{sec:Results}. Section \ref{sec:Discussion} concludes the paper while also giving an outlook on potential extensions of our work.

\section{Data}
\label{sec:data}

\begin{figure}[t]
\center
\includegraphics[width = 0.49\textwidth]{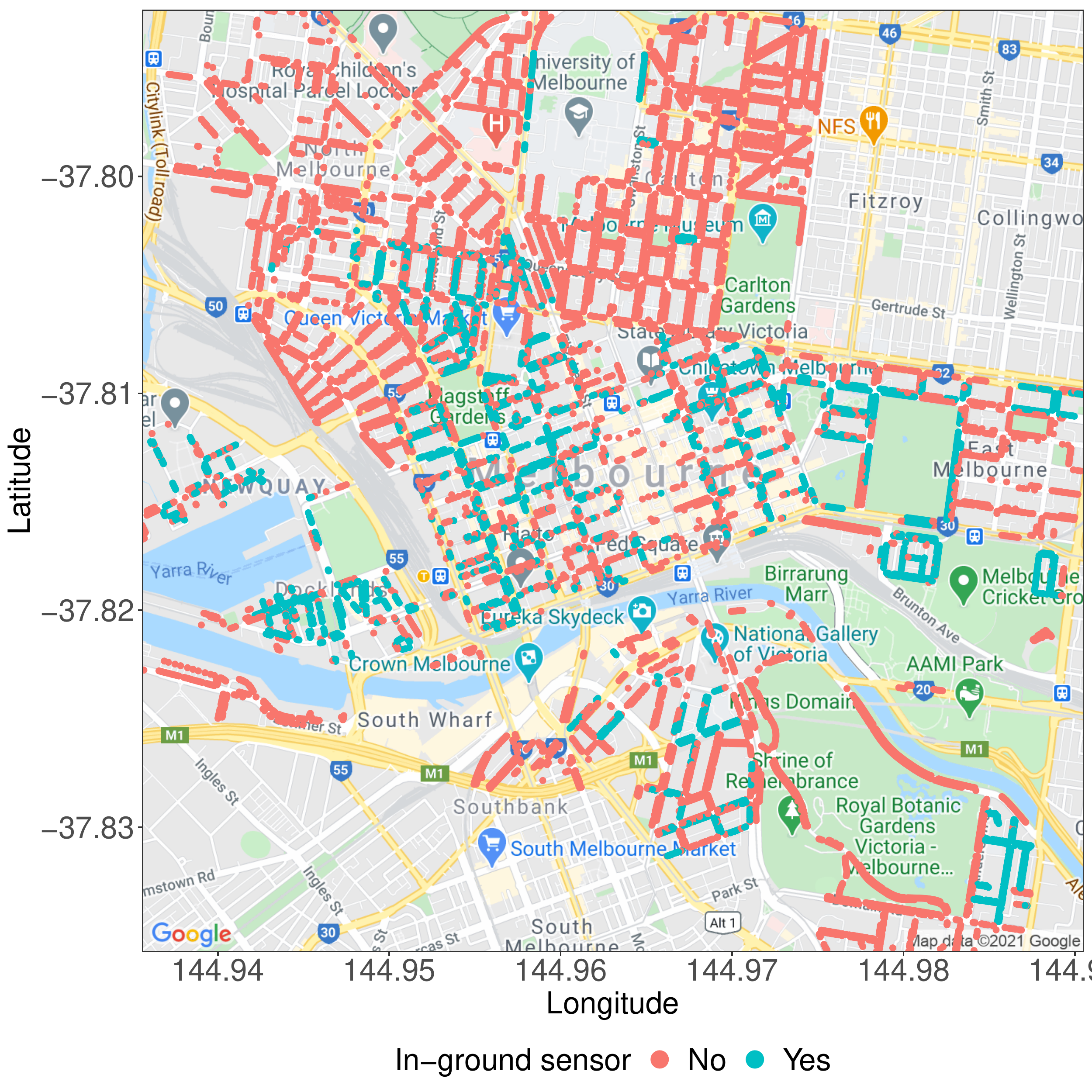} 
\caption{Location of on-street parking lots with and without in-ground sensors in the city center of Melbourne, Australia.}
\label{fig:sensors_yes_no}
\end{figure}

\begin{table}
\center
\resizebox{\textwidth}{!} {\begin{tabular}{cccccc}
\toprule
Start & End & Duration (minutes) & State & Marker & Side of street \\
\midrule
2019-04-30 08:24:11 & 2019-04-30 08:29:31 & 5.33 & 1 & 1155W & west \\
2019-04-30 08:29:31 & 2019-04-30 08:34:54 & 5.38 & 0 & 1155W & west \\
2019-04-30 08:34:54 & 2019-04-30 08:37:22 & 2.47 & 1 & 1155W & west \\
\vdots & \vdots & \vdots & \vdots & \vdots & \vdots \\
2019-06-07 08:53:44 & 2019-06-07 08:54:27 & 0.72 & 0 & C1170 & central \\
2019-06-07 08:54:27 & 2019-06-07 08:55:21 & 0.90 & 1 & C1170 & central \\
2019-06-07 09:15:11 & 2019-06-07 09:20:40 & 5.48 & 1 & C1170 & central \\
2019-06-07 09:20:40 & 2019-06-07 09:55:30 & 34.83 & 0 & C1170 & central \\
\bottomrule
\end{tabular}}
\caption{Structure of the preprocessed on-street parking lot data.}
\label{tab:data_structure}
\end{table}

We apply the model which we develop in this paper to on-street parking lot data from the City of Melbourne, Australia. The data originate from the year 2019 and are provided through the open data platform \cite{melbourne2020}. This database is filled by in-ground sensors which are installed underneath around 5,000 out of more than 20,000 on-street parking lots in the city center of Melbourne. Figure \ref{fig:sensors_yes_no} shows the location of these parking lots and we see that most of the sensors are located in the central business district (CBD) of Melbourne.\footnote{All maps in this paper are created using the R package \cite{ggmap2013}.} The basic structure of the already preprocessed data is exemplified in Table \ref{tab:data_structure}, where every row matches to a duration in either state 0 (clear) or 1 (occupied) that is specified to the second. The first three rows of Table \ref{tab:data_structure} correspond to consecutive events on the same parking lot, which can be identified through a unique marker. However, the last four rows show that the data is not complete, i.e. there are time intervals in which the sensor was either disabled or it just malfunctioned. Therefore, we advocate that these data are missing completely at random \citep{heitjan1996distinguishing}, i.e. not including them in the analyses does not lead to biased estimates. In other words, the available data can be considered as a random sample of the complete data. Altogether, the dataset for the year 2019 consists of more than 30 million observations in both states 0 and 1. 

\begin{figure}[t]
\center
\includegraphics[width = 0.49\textwidth]{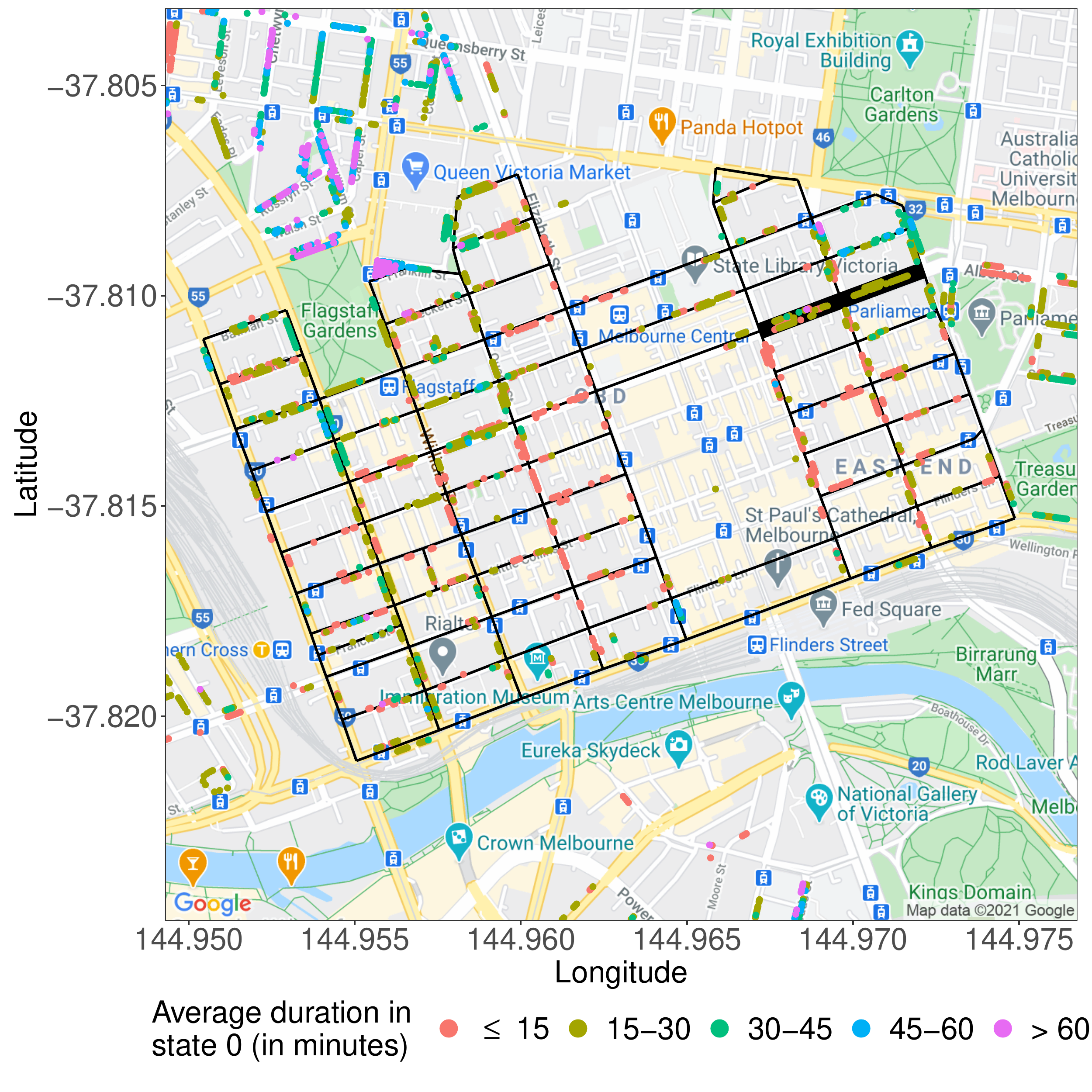} 
\includegraphics[width = 0.49\textwidth]{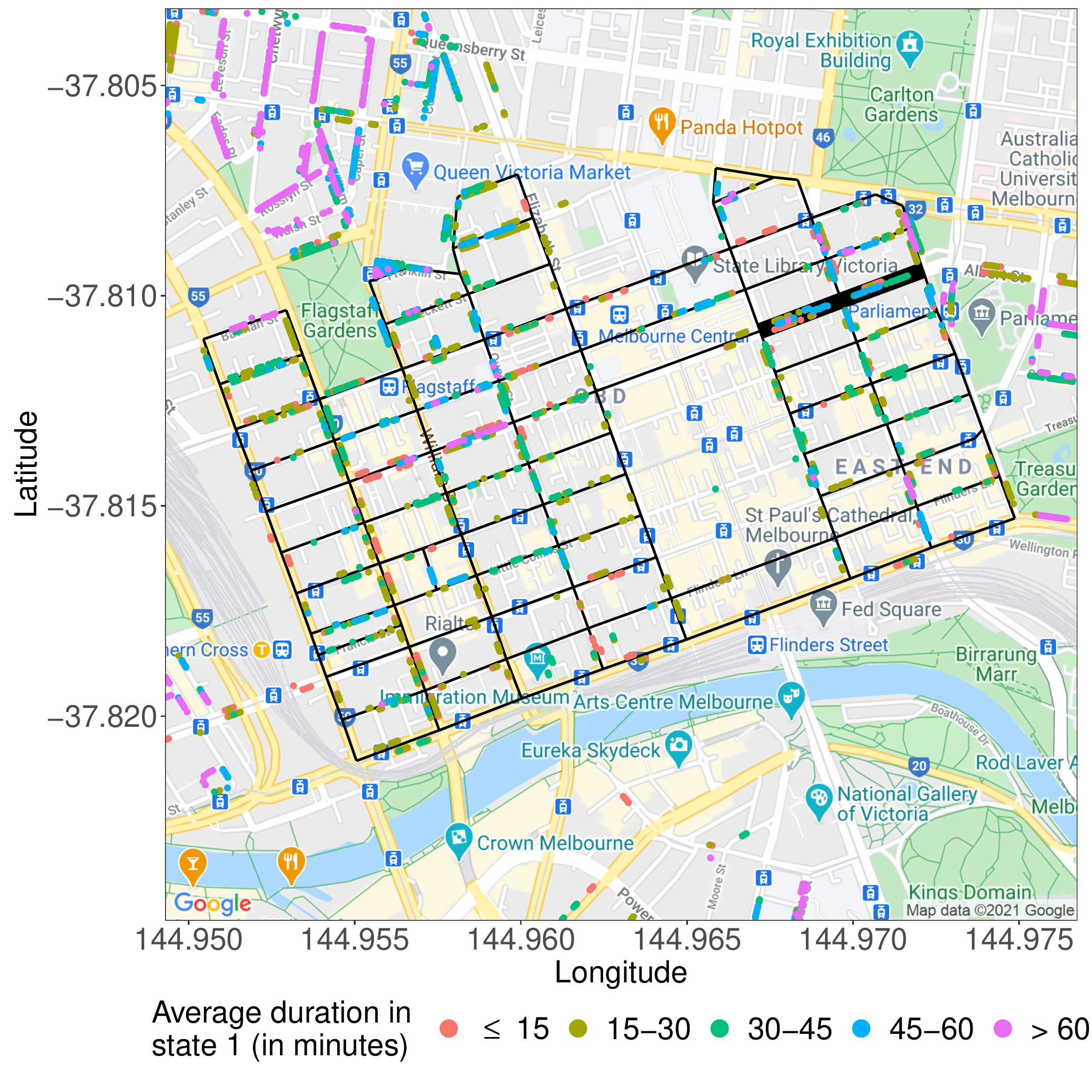}
\caption{Average duration of parking events in the CBD of Melbourne.}
\label{fig:average_duration}
\end{figure}

Figure \ref{fig:average_duration} shows parking lots in the CBD along with their average time being clear (state 0, left panel) or occupied (state 1, right panel), where we only considered parking events which started after 8 am and terminated before 8 pm on each day of the week. We see that parking lots around shopping malls in the center of the map tend to be unoccupied only for several minutes, whereas parking lots in the north-western part of the map extract are usually available for more than 30 minutes before they get occupied again. Here, the parking duration is usually more than one hour compared with usually less than 30 minutes in the center of the map extract. Note that the duration of parking is also affected by several parking restrictions which might differ even along the same street section. 

\begin{figure}[t]
\center 
\includegraphics[width = 0.49\textwidth]{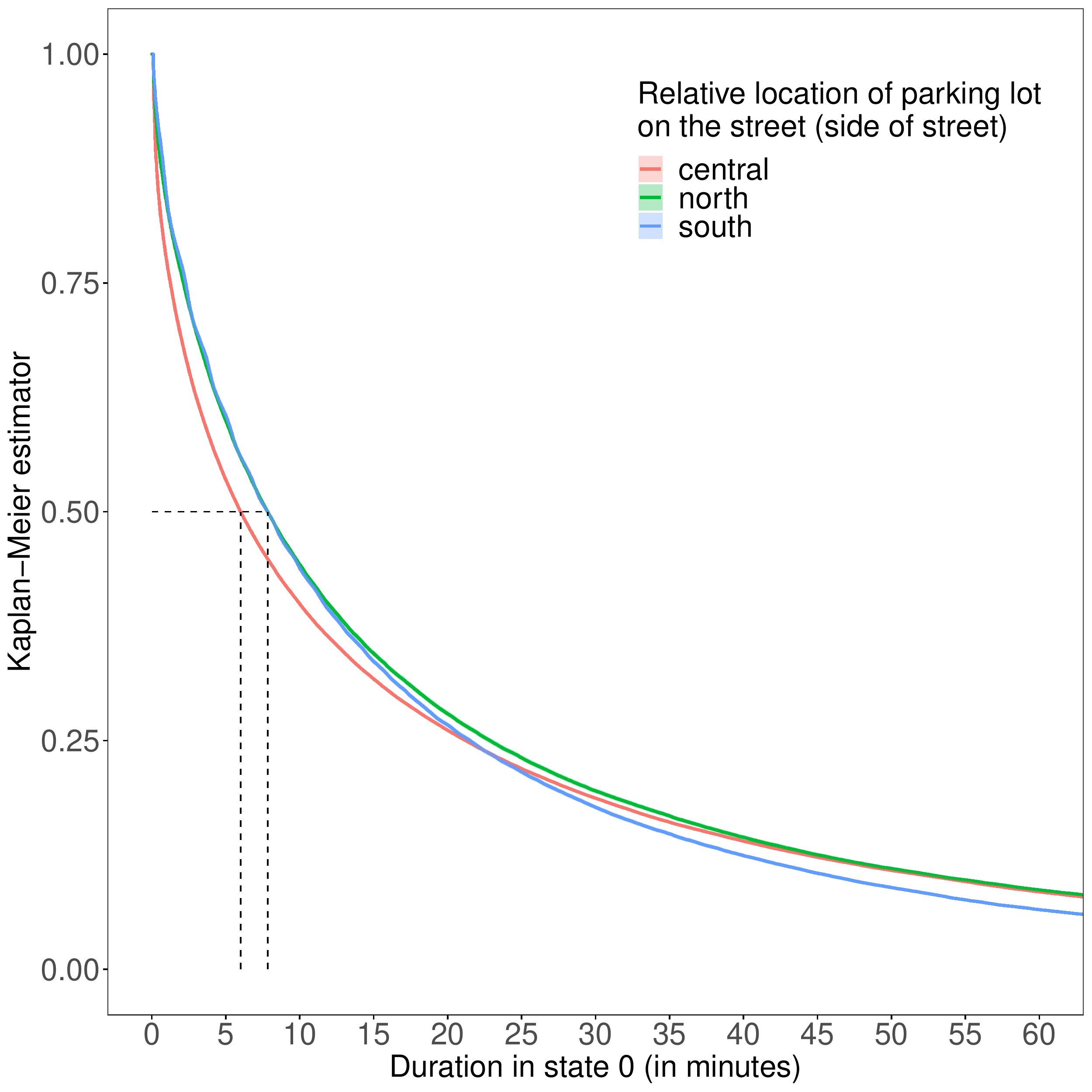} 
\includegraphics[width = 0.49\textwidth]{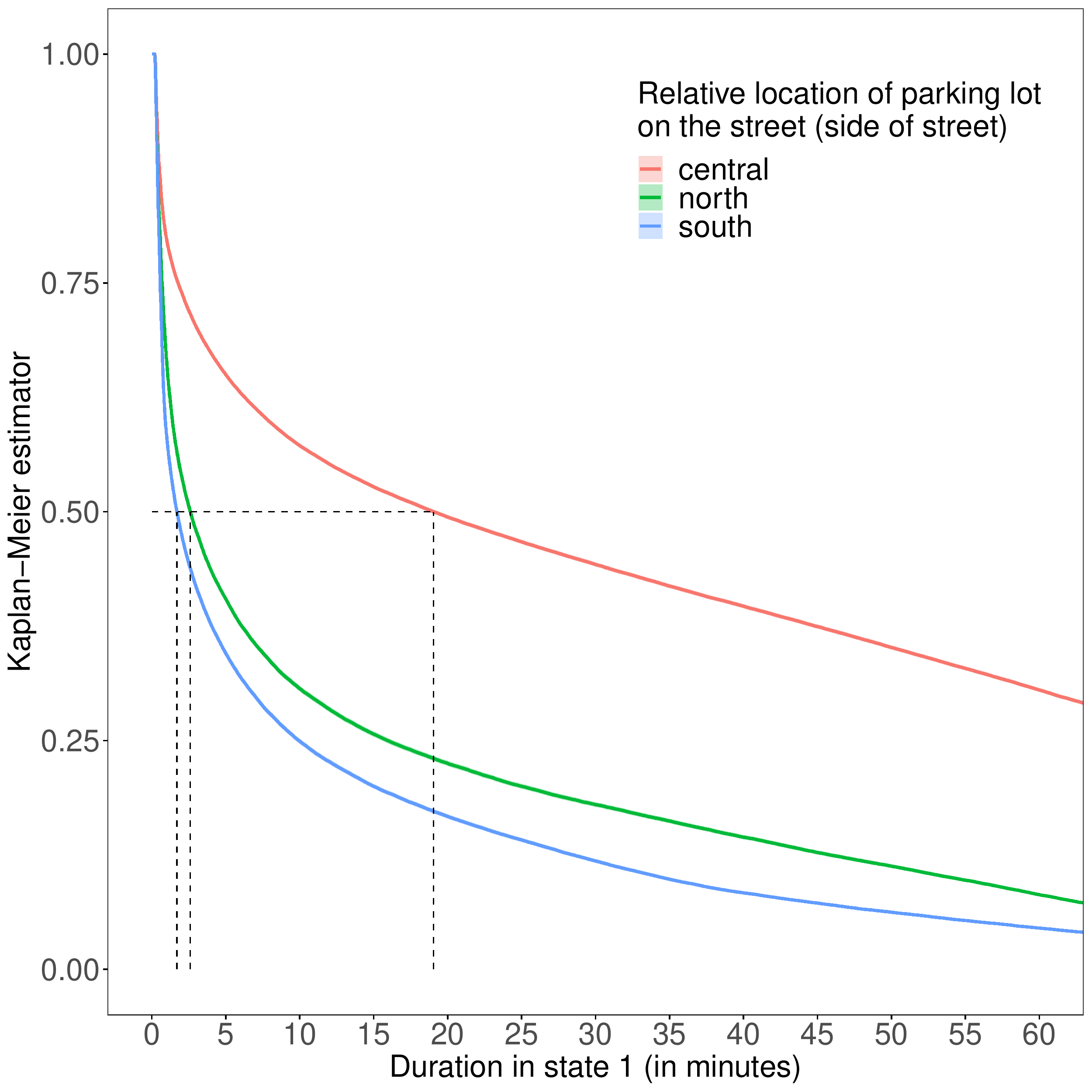}
\caption{Kaplan Meier estimators for durations in Lonsdale Street between Russel Street and Exhibition Street.}
\label{fig:KM_lonsdale_east}
\end{figure}

Looking very precisely at the plots in Figure \ref{fig:average_duration} it is already evident that the duration in both states 0 and 1 depends on the relative location of the parking lot on the street. In order to quantify this we show in Figure \ref{fig:KM_lonsdale_east} Kaplan-Meier estimators \citep{klein2006survival} of the duration in both states 0 and 1, restricted to one specific street segment which is highlighted as thicker lines in Figure \ref{fig:average_duration} (Lonsdale Street between Russel Street and Exhibition Street). We find that parking lots which are located in the center of this street segment, i.e. in between the two lanes, exhibit the longest median time being occupied. Furthermore, especially the Kaplan-Meier plot for state 1 suggests the duration times being not exponentially distributed. This already hints at the necessity of employing a  model which is capable of capturing also non-exponential duration times.

\section{Problem and notation}
\label{sec:notation}

We consider a set of on-street parking lots, indexed by $i = 1,\dots,N$, which are distributed along a network of streets, typically in the center of an urban area. A parking lot can be either clear or occupied which is why we model each parking lot as a continuous time two-state stochastic process $X^{(i)}$ with state space $\mathcal{S} = \lbrace 0,1 \rbrace$. In particular, $X_t^{(i)} = 0$ if parking lot $i$ is unoccupied at time point $t$ and $X_t^{(i)} = 1$ otherwise. We assume knowledge of the process $X^{(i)}$ from a time point $t_p < 0$ in the past until the present moment $t = 0$, where we additionally observe $K$ covariates $\boldsymbol{z}_t^{(i)} \in \mathbb{R}^K$ with $t \in [t_p,0]$ and  $i = 1,\dots,N$. Parking lots are located on a street network $\boldsymbol{G}$ which is why we consider a distance measure taking the geometry of this network into account \citep{baddeley2015spatial}. Defining with $\boldsymbol{s}_i \in \boldsymbol{G}$ the location of the $i$-th parking lot we obtain with $d_{\boldsymbol{G}}(\boldsymbol{s}_{i_1}, \boldsymbol{s}_{i_2})$ the (street-)network-based distance between the two parking lots indexed by $i_1$ and $i_2$, that is the driving distance between $\boldsymbol{s}_{i_1}$ and $\boldsymbol{s}_{i_2}$ with respect to $\boldsymbol{G}$. For simplicity we assume symmetry, i.e. one-way streets are ignored for now. 

All available information from $t_p$ up to the present moment $t = 0$ is formally contained in $\mathcal{F} = \sigma(X_t^{(i)}, \boldsymbol{z}_{t}^{(i)}; t \in [t_p,0], i = 1,\dots,N)$, where $\mathcal{F}$ denotes the observed history of the processes related to parking lot occupancy and covariate information including the present moment. Our goal is to predict the probability that a parking lot is unoccupied at a future time point $t_f > 0$. Since $X_0^{(i)}$ is included in the history $\mathcal{F}$, this probability is given by 
\begin{align*}
\mathbb{P}(X_{t_f}^{(i)} = 0 \mid \mathcal{F}) = P_{00}^{(i)}(t_f)\cdot \mathbbm{1}_{\lbrace 0 \rbrace}(X_{0}^{(i)}) + P_{10}^{(i)}(t_f)\cdot \mathbbm{1}_{\lbrace 1 \rbrace}(X_{0}^{(i)}),
\end{align*}
where 
\begin{align}
P_{jk}^{(i)}(t_f) &= \mathbb{P}(X_{t_f}^{(i)} = k \mid \mathcal{F}, X_0^{(i)} = j), \quad j,k = 0,1,
\label{eq:transition_probabilities}
\end{align}
are transition probabilites to be determined. In order to predict \eqref{eq:transition_probabilities}, we first need to charaterize the stochastic processes $X^{(i)}$ in more depth. Therefore, we denote with $D_{j,t}^{(i)}$ the random duration time that parking lot $i$ remains in state $j \in \mathcal{S}$ with index $t$ indicating some time point. Furthermore, we define with
\begin{align}
\lambda_{j,t}^{(i)}(d \mid \boldsymbol{z}_{t}^{(i)}) = \lim_{\Delta d \downarrow 0} \frac{\mathbb{P}(d \leq D_{j,t}^{(i)} < d + \Delta d \mid D_{j,t}^{(i)} \geq d, \boldsymbol{z}_{t}^{(i)})}{\Delta d}
\label{eq:hazard_general}
\end{align}
the transition intensity from state $j$ to state $1-j$ depending on the current duration $d$ in state $j$ and for  time point $t$. Note that in the context of time to event analysis, \eqref{eq:hazard_general} is usually denoted as the hazard rate and can thus be interpreted as the instantaneous rate at which a parking lot is changing its state. For fixed $t$ we obtain the relation between the hazard function and the distribution function of the duration $D_{j,t}^{(i)}$ (see e.g. \citealp{kalbfleisch2011statistical})
\begin{align}
\mathbb{P}(D_{j,t}^{(i)} \leq d) = 1- \exp \left(- \int_0^d \lambda_{j,t}^{(i)}(x\mid \boldsymbol{z}_{t}^{(i)}) \dif x \right),
\label{eq:distribution_D}
\end{align}
where the integral in \eqref{eq:distribution_D} represents the cumulative hazard for duration $d$. It follows immediately that constant transition intensities imply exponentially distributed duration times $D$, i.e. the memorylessness property $\mathbb{P}(D > d+s \mid D > d) = \mathbb{P}(D > s)$ holds for $s,d > 0$. Allowing the transition intensities $\lambda(\cdot)$ to also depend on the  duration time $d$ in the current state leads to non-exponentially distributed duration times in general. We derive in the next section the transition probabilities \eqref{eq:transition_probabilities}  for the above defined process.


\section{Prediction of transition probabilities}
\label{sec:semiMarkov}


\subsection{Transition probabilities in semi-Markov processes}
 \label{subsec:semi_Markov}



Though our process has just two states, we provide a general description with multiple states here. We will see that this is advantageous as it will allow us to incorporate information about the duration from the most previous change of state before $t = 0$. More details are given in the next subsection. We define with $0 = t_{(0)} < t_{(1)} < t_{(2)} < {\dots}$ the time points of status changes of the stochastic process $X = (X_t)_{t \geq 0}$ whose finite state space is denoted as $\mathcal{S}$. Bear in mind that we consider the current time point as $t=0$. Further, we denote with $X_{t_{(n)}} \in \mathcal{S}$ the state of the process at the time of the $n$-th transition in which $X$ stays for the duration $D_{(n)}$. In other words, $D_{(n)}$ is the random length of the interval $[t_{(n)}, t_{(n+1)})$. We assume that $X$ is a semi-Markov process which is fully characterized by an initial distribution $\boldsymbol{p} = [p_j(0) \mid j \in S]$ with $\sum_{j \in \mathcal{S}} p_j(0) = 1$ and the renewal kernel $\boldsymbol{Q}(d) = [\mathcal{Q}_{jk}(d) \mid j,k \in \mathcal{S}]$, where 
\begin{align}
\begin{split}
\label{eq: Q-matrix}
\mathcal{Q}_{jk}(d) &= \mathbb{P}(X_{t_{(n+1)}} = k, D_{(n)} \leq d \mid X_{t_{(n)}} = j, D_{(n-1)}, X_{t_{(n-1)}}, \dots) \\
&=  \mathbb{P}(X_{t_{(n+1)}} = k, D_{(n)} \leq d \mid X_{t_{(n)}} = j)
\end{split}
\end{align}
for $d \geq 0$. Note that $X$ has right-continuous sample paths and by definition we have $\mathcal{Q}_{jj}(\cdot) \equiv 0$ for $j \in \mathcal{S}$. The cumulative conditional distribution function of $D_{(n)}$ is given by  $F_j(d)  = \mathbb{P}(D_{(n)} \leq d \mid X_{t_{(n)}} = j) = \sum_{k \in \mathcal{S}} \mathcal{Q}_{jk}(d)$ for $j \in \mathcal{S}$ and $n  \in \mathbb{N}_0$. 

From \eqref{eq:distribution_D} it follows that in case of non-constant transition intensities, $D_{(n)}$ satisfies the memorylessness property only in the instant $t_{(n)}$ of the transition into state $X_{t_{(n)}}$. Under consideration of this property it can be shown that for $t \geq 0$ the interval transition probabilities $P_{jk}(t) = \mathbb{P}(X_t = k \mid X_0 = j)$ are the solutions of the following integral equations
\begin{align}
\label{eq: integral equations}
P_{jk}(t) = (1- F_j(t))\delta_{jk} + \sum_{m \in \mathcal{S}} \int_0^t  P_{mk}(t-x)q_{jm}(x) \dif x 
\end{align}
with initial condition $P_{jk}(0) = \delta_{jk}$  (Kronecker delta) and $q_{jk}(\cdot)$ denotes the derivative of $\mathcal{Q}_{jk}(\cdot)$ with respect to the duration time. Note that if the subsequent state of each state $j \in \mathcal{S}$ is deterministic, say $k \in \mathcal{S}$, then $Q_{jk}(\cdot)$ is a probability distribution function and with $f_j(\cdot) = q_{jk}(\cdot)$ we denote the corresponding density. This will simplify matters later when we are specifically concerned with the parking lot problem. 

\begin{figure}
\center
\includegraphics[width = 0.6\textwidth]{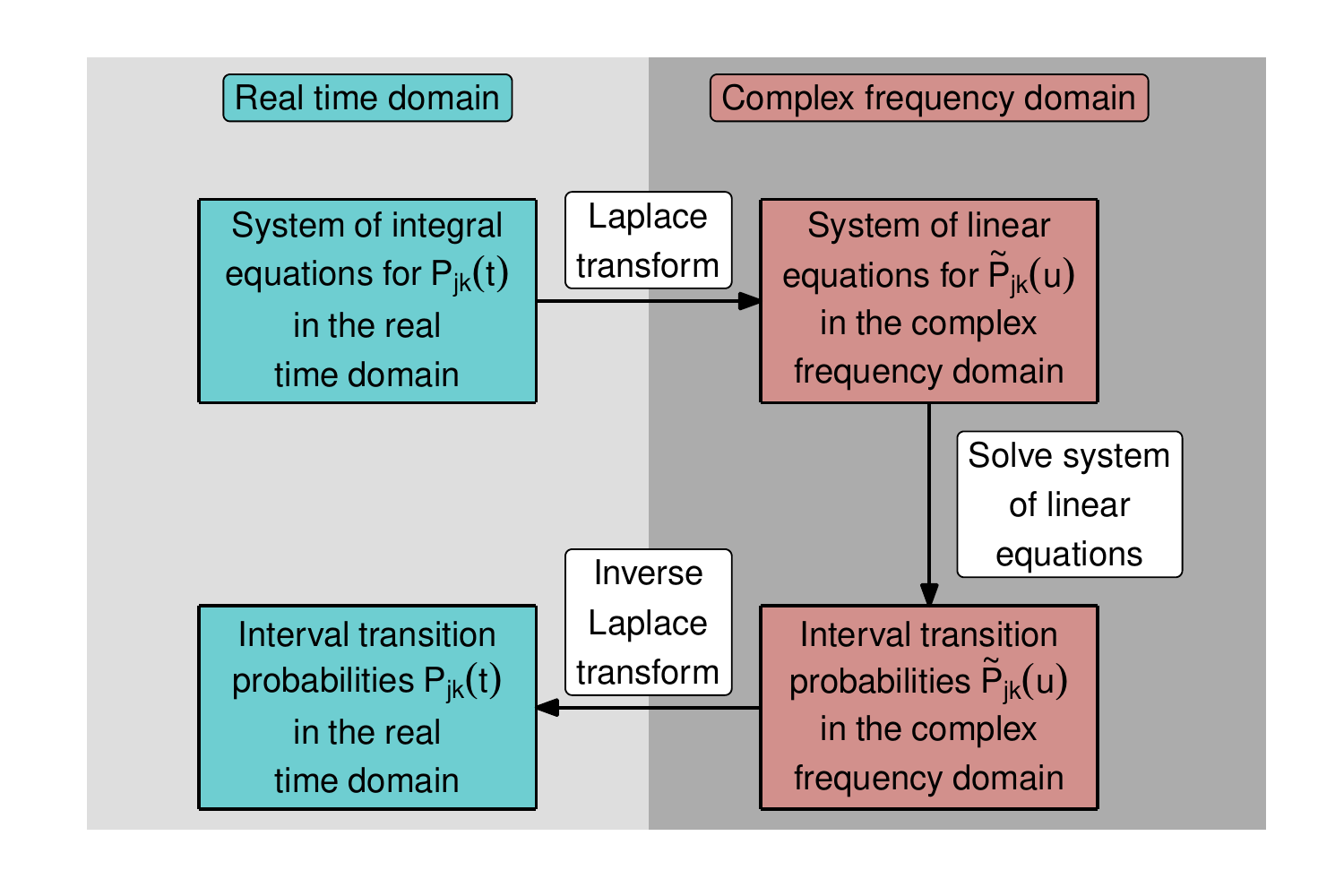}	
\caption{Visualization of the procedure of obtaining the interval transition probabilities $P_{jk}(t)$ in a semi-Markov process from the system of integral equations \eqref{eq: integral equations}.}
\label{fig:VisualizeLaPlace}
\end{figure}

Following \cite{grabski2014semi}, the systems of integral equations \eqref{eq: integral equations} can be solved via Laplace transforms (\citeauthor{widder2015laplace}, \citeyear{widder2015laplace}). The Laplace transform $\widetilde{f} := \mathcal{L}\lbrace f \rbrace: \mathbb{C} \supset C \rightarrow \mathbb{C}$ of a real valued function $f:[0, \infty) \rightarrow \mathbb{R}$ is given by
\begin{align}
\widetilde{f}(u) := \mathcal{L}\lbrace f \rbrace(u) = \int_0^\infty f(t) e^{-ut}dt,
\label{eq:laplacetrafo}
\end{align} 
where the integral in  \eqref{eq:laplacetrafo} converges for $u \in C$. The set $C$ is called the region of convergence and consists of all $u \in \mathbb{C}$ which satisfy $\mathfrak{Re}(u) > \gamma$, the so-called abscissa of convergence \citep{hall1992abscissa}, and $\mathfrak{Re}(u)$ denotes the real part of $u$. 
In particular, it holds that $\mathcal{L}\lbrace 1 \rbrace(u) = \frac{1}{u}$ and the Laplace transform of $\int_{0}^t P_{mk}(t-x) q_{jm}(x) \dif x$ is given by $\widetilde{P}_{mk}(u)\widetilde{q}_{jm}(u)$, i.e. convolution in the real time domain corresponds to multiplication in the complex frequency domain. Consequently, the linearity of the Laplace transform easily allows to represent the system of integral equations \eqref{eq: integral equations} as the following system of linear equations in the frequency domain
\begin{equation}
\label{eq: transformed equations}
\widetilde{P}_{jk}(u) = \left( \frac{1}{u} - \widetilde{F}_j(u) \right) \delta_{jk} + \sum_{m \in \mathcal{S}} \widetilde{q}_{jm}(u) \widetilde{P}_{mk}(u).
\end{equation}
Defining with $\widetilde{\boldsymbol{q}}(u) = [\widetilde{q}_{jk}(u) \mid j,k \in \mathcal{S}]$ and $\widetilde{\boldsymbol{F}}(u) = [\delta_{jk}\widetilde{F}_j (u)\mid j,k \in \mathcal{S}]$ matrices of the same dimension as $\boldsymbol{Q}(u)$, we obtain the solution of \eqref{eq: transformed equations} as
\begin{align}
\widetilde{\boldsymbol{P}}(u) = \left(\boldsymbol{I}-\widetilde{\boldsymbol{q}}(u)\right)^{-1} \left( \frac{1}{u} \boldsymbol{I} - \widetilde{\boldsymbol{F}}(u) \right),
\label{eq:solution_lin}
\end{align}
where $\boldsymbol{I}$ denotes the identity matrix and $\widetilde{\boldsymbol{P}}(u) =  [\widetilde{P}_{jk}(u) \mid j,k \in \mathcal{S}]$. The interval transition probabilities $P_{jk}(t)$ in the real time domain can then be calculated by applying the inverse Laplace transform element-wise to the solution \eqref{eq:solution_lin} in the complex frequency domain. The inverse Laplace transform $\mathcal{L}^{-1}$ of a Laplace-transformed function $\mathcal{L}\lbrace f \rbrace : \mathbb{C} \rightarrow \mathbb{C}$ is generally defined through the following Bromwich integral \citep{weideman2007parabolic}
\begin{align*}
f(t) = \mathcal{L}^{-1} \lbrace \mathcal{L} \lbrace f \rbrace(u) \rbrace (t) = \frac{1}{2\pi i} \lim_{T \rightarrow \infty} \int_{\gamma_0 - iT}^{\gamma_0 + iT} e^{ut} \mathcal{L}\lbrace f \rbrace (u) \dif u,
\end{align*}
where $\gamma_0 > \gamma$ from above, i.e. $\gamma_0$ must be in the region of convergence $C$ of $\mathcal{L}\lbrace	f \rbrace$. The whole procedure for determining the interval transition probabilities \eqref{eq:transition_probabilities} which we formally described above is summarized in Figure \ref{fig:VisualizeLaPlace}.

\subsection{Solving the parking lot problem}
\label{subsec:semiMarkov_parking}

Suppose that we find ourselves at the current time point $t = 0$ and for simplicity of notation, we drop superscript $(i)$ related to the parking lot index in this section. In order to predict the transition probabilities \eqref{eq:transition_probabilities} in the time span ranging from $t = 0$ until a future time point $t_f>0$ we need $\lambda_{j,t}(d)$ for $t > 0$. Certainly, the future evolution of time dependent covariates is unknown at $t = 0$, i.e. $\lambda_{j,t}(d)$ might not be available for $t > 0$. However, since in our specific application to parking lot data the forecast horizon is usually in the range of several minutes up to an hour, we assume that $\lambda_{j,t}(d) = \lambda_{j,t=0}(d)$ which we denote in short with $\lambda_{j}(d) = \lambda_{j,t=0}(d)$ for $j = 0,1$. Estimation of $\lambda_j(\cdot)$ is covered in the following Section \ref{sec:intensities}. 

\begin{figure}
\center
\includegraphics[width = \textwidth]{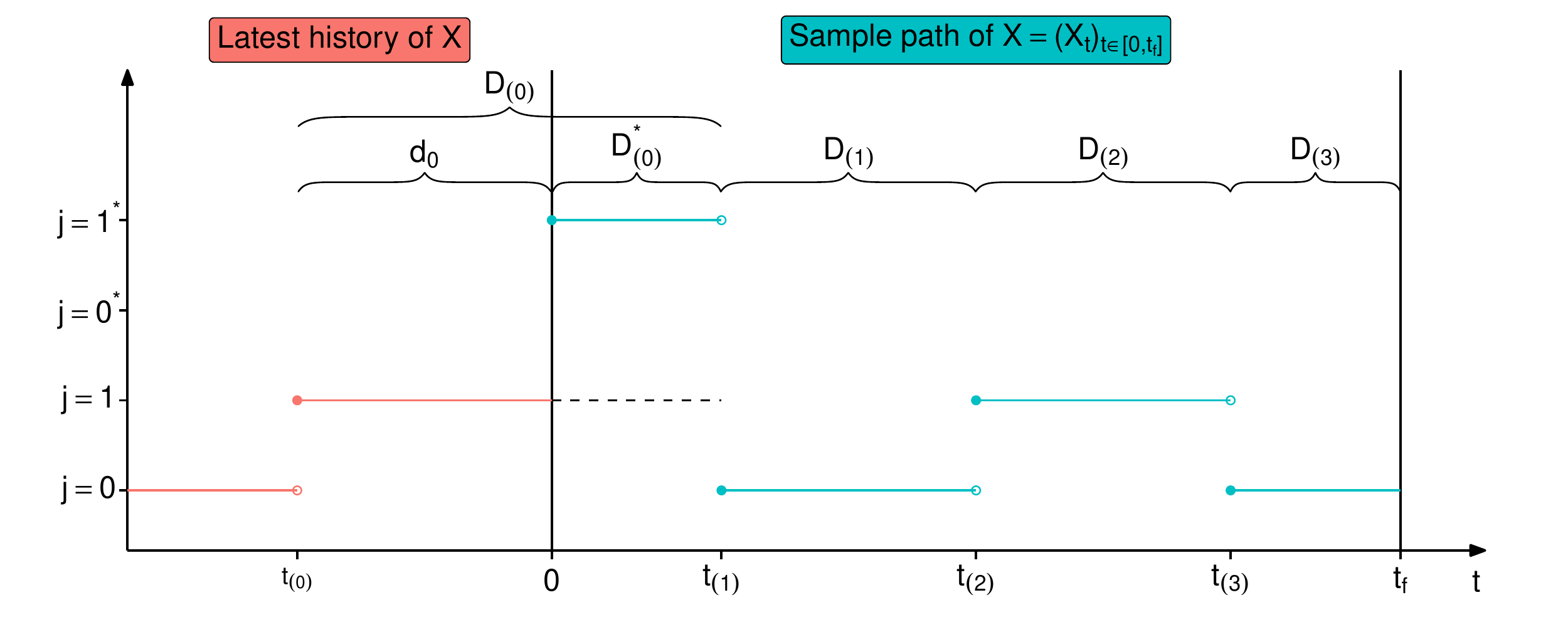}	
\caption{Visualization of the process $X = (X_t)_{t \in [0,t_f]}$ in view of the current time point $t = 0$.}
\label{fig:VisualizeSemiMarkov}
\end{figure}

We can now adopt the theoretical concepts outlined in Section \ref{subsec:semi_Markov} to our problem. First, the history $\mathcal{F}$ contains the state of each parking lot at the present moment $t = 0$, i.e. the initial distribution $\boldsymbol{p}$ is deterministic. However, formula \eqref{eq: integral equations} derived above is based on the assumption that the process $X$ jumps in $t = 0$ into its initial state, i.e. $t_{(0)} = 0$. In other words, the duration in the initial state at $t = 0$ is zero. This is certainly not the case with regards to our parking lot problem since at $t = 0$ a parking lot has already been in state $ X_0 = j$ for a known duration, say $d_{0}$. This is shown in Figure \ref{fig:VisualizeSemiMarkov} where, in view of the present moment $t = 0$, the parking lot has lastly changed its state at time point $t_{(0)} < 0$. Therefore, for the random duration time $D_{(0)}^\star = t_{(1)}$ from being in state $j$ at $t = 0$ until the first transition to state $1-j$ it holds that
\begin{align}
\resizebox{.9 \textwidth}{!} {$\mathbb{P}(D_{(0)}^\star > d \mid X_0 = j) = \mathbb{P}(D_{(0)} > d + d_{0} \mid D_{(0)} > d_{0}, X_0 = j) = \exp\left( - \int_{d_{0}}^{d_{0}+d} \lambda_{j}(x)\dif x \right)$} .
\label{eq:dist_D_star}
\end{align} 
Consequently, if $t_{(0)} < 0$ the distribution of the duration $D_{(0)}^\star$ in the initial state differs from the distribution of the duration $D_{(0)} = t_{(1)} - t_{(0)}$. This is illustrated in Figure \ref{fig:VisualizeSemiMarkov} where it holds for the duration times in the occupied state that $D_{(0)}^\star \overset{d}{\neq} D_{(k)} = t_{(k+1)} - t_{(k)}$ for $k = 2, 4, 6,\dots$. In order to respect this finding in our model we add two initial states $0^\star$ and $1^\star$ to the updated state space $\mathcal{S}^\star = \lbrace 0^\star, 1^\star, 0,1 \rbrace$, where $j^\star$ is the state of $X$ at $t = 0$. 

Further, it holds that $F_{j^\star}(\cdot) = \mathcal{Q}_{j^\star,1-j}(\cdot)$ and $F_{j}(\cdot) = \mathcal{Q}_{j,1-j}(\cdot)$ for $j = 0,1$ with
\begin{align*}
F_{j^\star}(d) = 1-\exp\left(- \int_{d_{0}}^{d_{0}+d} \lambda_{j}(x) \dif x \right) \quad\text{and}\quad F_j(d) = 1-\exp\left(- \int_0^d \lambda_{j}(x) \dif x \right).
\end{align*}
With $f_{j^\star}(\cdot) = q_{j^\star,1-j}(\cdot)$ and $f_{j}(\cdot) = q_{j,1-j}(\cdot)$ we denote the densities corresponding to the distribution functions $F_{j^\star}$ and $F_j$, respectively. We can now employ the general solution \eqref{eq:solution_lin} in the setting with state space $\mathcal{S}^\star = \lbrace 0^\star, 1^\star, 0, 1\rbrace$  which yields a matrix $\widetilde{\boldsymbol{P}}(u) = [\widetilde{P}_{jk} \mid j,k \in \mathcal{S}^\star] \in \mathbb{C}^{4 \times 4}$  of interval transition probabilities  represented in the complex frequency domain. The matrix $\boldsymbol{I}-\widetilde{\boldsymbol{q}}(u)$ in \eqref{eq:solution_lin} and its inverse are given by
\begin{align*}
\resizebox{.9 \textwidth}{!} {$ \boldsymbol{I} - \widetilde{\boldsymbol{q}}(u) = \begin{pmatrix}
1 & 0 & 0 & -\widetilde{f}_{0^\star}(u) \\
0 & 1 & -\widetilde{f}_{1^\star}(u) & 0 \\
0 & 0 & 1 & -\widetilde{f}_{0}(u) \\
0 & 0 & -\widetilde{f}_{1}(u) & 1
\end{pmatrix}, \quad
 (\boldsymbol{I} - \widetilde{\boldsymbol{q}}(u))^{-1} = 
\begin{pmatrix}
1 & 0 & \frac{\widetilde{f}_{0^\star}(u)\widetilde{f}_{1}(u)}{1-\widetilde{f}_{0}(u)\widetilde{f}_{1}(u)} & \frac{\widetilde{f}_{0^\star}(u)}{1-\widetilde{f}_{0}(u)\widetilde{f}_{1}(u)} \\
0 & 1 & \frac{\widetilde{f}_{1^\star}(u)}{1-\widetilde{f}_{0}(u)\widetilde{f}_{1}(u)} & \frac{\widetilde{f}_{1^\star}(u)\widetilde{f}_{0}(u)}{1-\widetilde{f}_{0}(u)\widetilde{f}_{1}(u)} \\
0 & 0 & 1 &  \frac{\widetilde{f}_{0}(u)}{1-\widetilde{f}_{0}(u)\widetilde{f}_{1}(u)}\\
0 & 0 & \frac{\widetilde{f}_{1}(u)}{1-\widetilde{f}_{0}(u)\widetilde{f}_{1}(u)} & 1
\end{pmatrix} $}
\end{align*}
respectively. Now, using formula \eqref{eq:solution_lin} and denoting $\widetilde{f}(u) = (1-\widetilde{f}_0(u)\widetilde{f}_1(u))^{-1}$ finally yields the solution of \eqref{eq: transformed equations} as
\begin{align}
\resizebox{.9 \textwidth}{!} {$\widetilde{\boldsymbol{P}}(u) 
=
 \begin{blockarray}{ccccc}
 & \colorbox{blue!30}{$0^\star$} & \colorbox{red!30}{$1^\star$} & \colorbox{blue!30}{0 } & \colorbox{red!30}{1 } \\ 
 \begin{block}{c(cccc)}
\colorbox{blue!30}{$0^\star$} &  \frac{1}{u} - \widetilde{F}_{0^\star}(u) & 0 & \widetilde{f}(u)\widetilde{f}_{0^\star}(u)\widetilde{f}_{1}(u)\left( \frac{1}{u} - \widetilde{F}_0(u) \right) &  
 \widetilde{f}(u)\widetilde{f}_{0^\star }(u)\left( \frac{1}{u} - \widetilde{F}_1(u) \right) \\
\colorbox{red!30}{$1^\star$} & 0 & \frac{1}{u} - \widetilde{F}_{1^\star}(u) & 
 \widetilde{f}(u)\widetilde{f}_{1^\star }(u)\left( \frac{1}{u} - \widetilde{F}_0(u) \right) & 
 \widetilde{f}(u)\widetilde{f}_{1^\star }(u)\widetilde{f}_{0}(u)\left( \frac{1}{u} - \widetilde{F}_1(u) \right) \\
\colorbox{blue!30}{0 } &  0 & 0 & \widetilde{f}(u)\left(\frac{1}{u} - \widetilde{F}_0(u)\right) & \widetilde{f}(u)\widetilde{f}_{0}(u) \left( \frac{1}{u} - \widetilde{F}_1(u) \right) \\
\colorbox{red!30}{1 } & 0 & 0 & \widetilde{f}(u)\widetilde{f}_{1}(u) \left( \frac{1}{u} - \widetilde{F}_0(u) \right) & \widetilde{f}(u)\left(\frac{1}{u} - \widetilde{F}_1(u)\right) \\
 \end{block}
\end{blockarray}$}.
 \label{eq:solution_complex}
\end{align}
The interval transition probabilities $P_{jk}(t)$ can be obtained by applying the inverse Laplace transform to the entries of $\widetilde{\boldsymbol{P}}(u)$, i.e. $P_{jk}(t) = \mathcal{L}^{-1} \lbrace \widetilde{P}_{jk}(u)\rbrace(t)$ for $j,k \in \mathcal{S}^\star$ and $t \geq  0$. Since $\mathcal{L}^{-1}\lbrace 1/u\rbrace(t) = \mathbbm{1}_{[0,\infty)}(t)$, it particularly holds that $P_{0^\star0^\star}(t) = 1-F_{0^\star}(t)$ and $P_{1^\star1^\star}(t) = 1-F_{1^\star}(t)$ for $t \geq 0$. This is an expected result since $P_{j^\star j^\star}(t)$ is the probability that a  parking lot's state will remain unchanged from time point $t = 0$ until time $t > 0$ and $1-F_{j^\star}(t) = \mathbb{P}(D_{(0)}^\star > t \mid X_0 = j)$ represents the probability that the duration in the initial state $j$ will exceed time point $t$. For the remaining entries of the transition probability matrix $\boldsymbol{P}(t)$ we have to build on numerical inversion techniques here, where we make use of the approach proposed by \cite{valsa1998approximate}. Details about the algorithm can be found in Appendix \ref{app:invlap}. In the end, the probability that a parking lot is clear at a prospective time point $t_f > 0$, conditional on the history $\mathcal{F}$ of parking lot occupation and covariate information, is given by
\begin{align}
\mathbb{P}(X_{t_f} = 0 \mid \mathcal{F}) = \begin{cases}
1 - F_{0^\star}(t) + \mathcal{L}^{-1}\lbrace\widetilde{P}_{0^\star 0}(u)\rbrace(t_f), & X_{0} = 0 \\
\mathcal{L}^{-1}\lbrace\widetilde{P}_{1^\star 0}(u)\rbrace(t_f), & X_{0} = 1
\end{cases}.
\label{eq:prediction_semiMarkov}
\end{align}

\subsection{Exponentially distributed duration times}
\label{subsec:exp}

Using the techniques from above, explicit formulae can be derived for the transition probabilities \eqref{eq:transition_probabilities} if the durations $D_{(n)}$ are exponentially distributed, i.e. if $X = (X_t)_{t \in [0,t_f]}$ is in fact a Markov process. With the transition intensities $\lambda_j$ being constant, it holds that $F_{j^\star}(d) = F_j(d) = 1 - e^{-\lambda_jd}$ and $f_j(d) = \lambda_je^{-\lambda_jd}$. Therefore, we can omit the states $0^\star$ and $1^\star$ and for the interval transition probabilities it follows that 
\begin{equation}
\label{eq: P two state Markov}
\boldsymbol{P}(t) = \frac{1}{\lambda_{0} + \lambda_{1}} \begin{pmatrix}
\lambda_{1} + \lambda_{0} e^{-t(\lambda_{0} + \lambda_{1})} &
\lambda_{0}\left[ 1- e^{-t(\lambda_{0} + \lambda_{1})} \right]  \\
\lambda_{1}\left[ 1- e^{-t(\lambda_{0} + \lambda_{1})} \right]  & 
\lambda_{0} + \lambda_{1} e^{-t(\lambda_{0} + \lambda_{1})}
\end{pmatrix}.
\end{equation}
Usually, the result \eqref{eq: P two state Markov} is derived by solving the Kolomogorov forward differential equations as e.g. in \cite{ross1996stochastic}, Example 5.4(A). 

\section{Estimation of transition intensities}
\label{sec:intensities}

\subsection{Modeling transition intensities}

We now discuss the estimation of transition intensities $\lambda_{j,t}^{(i)}(d \mid \boldsymbol{z}_t^{(i)})$ as defined in \eqref{eq:hazard_general} which can be interpreted as hazard rates in a time to event model. We consider covariates $\boldsymbol{z}_t^{(i)} = (z_{1,t}^{(i)},\dots,z_{K,t}^{(i)})^\top$ through a model of the form 
\begin{align}
\lambda_{j,t}^{(i)}(d  \mid \boldsymbol{z}_{t}^{(i)}) &= \lambda_{j,0}(d) \exp\left(\beta_{0,j}+\sum_{k=1}^K g_{j,k}(z_{k,t}^{(i)}) + u_{j}^{(i)}\right).
\label{eq:model}
\end{align}
Here, $\lambda_{j,0}(\cdot)$ is the common baseline intensity for the transition from state $j$ to state $1-j$. Furthermore, $\eta_{j,t}^{(i)}= \beta_{0,j} + \sum_{k=1}^K g_{j,k}(z_{k,t}^{(i)})$ is the linear predictor of the model (including the intercept $\beta_{0,j}$) which will be treated in more depth later. Coefficients $u_{j}^{(i)}$ are random effects, which account for unobserved parking lot specific heterogeneity. Hereby, we assume that $v_j^{(i)} = \log u_j^{(i)}$ follow indenpendently a $\text{Gamma}(\frac{1}{\gamma_j}, \frac{1}{\gamma_j})$ prior distribution with $\mathbb{E}(v_j^{(i)}) = 1$ and $\text{Var}(v_j^{(i)}) = \gamma_j$. In the context of time to event analysis, these kinds of models are known as (gamma) shared frailty models \citep{therneau2003penalized}, since a common multiplicative frailty $v_j^{(i)}$ on the baseline hazard is shared among observations for parking lot $i$  in state $j$.

\subsection{Choosing an appropriate baseline intensity}

Under consideration of \eqref{eq:distribution_D} and \eqref{eq:model}, the distribution function $F_j^{(i)}(\cdot)$ of the random duration $D_{j}^{(i)}$ that parking lot $i$ stays in state $j$ is given by
\begin{align}
\label{eq:distribution_function}
F_{j}^{(i)}(d) = \mathbb{P}(D_{j}^{(i)} \leq d) =  1- \exp \left( - \exp(\eta_{j,t=0}^{(i)} + u_j^{(i)}) \int_0^d \lambda_{j,0}(x) \dif x \right)
\end{align}
for $j = 0,1$. For $j = 0^\star,1^\star$ the integral boundaries in \eqref{eq:distribution_function} need to be shifted by the current duration $d_0$ as motivated above. Employing the numerical algorithm proposed by \cite{valsa1998approximate} in order to compute the inverse Laplace transformation of \eqref{eq:solution_complex} involves repeatedly evaluation of both $F_j^{(i)}(\cdot)$ and its derivative $f_j^{(i)}(\cdot)$. Considering \eqref{eq:distribution_function}, it is evident that the form of the baseline intensity $\lambda_{j,0}(\cdot)$ determines the numerical effort therefore, which is why an easy-to-integrate $\lambda_{j,0}(\cdot)$ is preferred. Often, the  baseline intensity is not explicitely modeled as in the Cox-Model. Here, the (cumulative) baseline intensity can be obtained via the Breslow estimator \citep{lin2007breslow}. Alternatively, the baseline can be modeled semiparametrically, e.g. with piece-wise exponential additive mixed models (PAMMs, \citealp{bender2018generalized}). However, with both approaches the cumulative hazard can only be evaluated numerically which has the consequence that the  inversion of the Laplace transform suffers from numerical instability. The employment of a fully-parametric model for $\lambda_{j,0}(\cdot)$ allows to circumvent numerical integration if the integral of $\lambda_{j,0}(\cdot)$ has an explicit representation. This is pursued in the following. 

A frequently used parametric time to event model is the Weibull model, in which case \eqref{eq:model} for $t = 0$ has the form $\lambda_{j}^{(i)}(d) = b_{j}^{(i)}\alpha_jd^{\alpha_j-1}$, where $\alpha_jd^{\alpha_j-1} = \lambda_{j,0}(d; \alpha_j)$ is the common baseline intensity and $b_{j}^{(i)} = \exp(\eta_{j,t=0}^{(i)} + u_j^{(i)})$.  This allows to express both the distribution function $F_j^{(i)}(d) = 1 - \exp \left( -b_{j}^{(i)}d^{\alpha_j}\right)$ and density $f_j^{(i)}(d) = b_{j}^{(i)} \alpha_j d^{\alpha_j-1} \exp \left( -b_{j}^{(i)}d^{\alpha_j}\right)$ for $d \geq 0$ explicitly. Finally, note that the baseline intensity $\lambda_{j,0}(\cdot)$ is shaped by a single parameter $\alpha_j$, with $\alpha_j < 1$ $(\alpha_j > 1)$ resulting in strictly decreasing (increasing) transition intensities. 

\subsection{The linear predictor}

The linear predictor $\eta_{j,t}^{(i)}= \beta_{0,j} + \sum_{k=1}^K g_{k,j}(z_{k,t}^{(i)})$ is independent of the duration time $d$ and multiplicatively takes the covariate effects in the time to event model \eqref{eq:model} into account. We either include the $k$-th covariate linearly in the model, i.e. $g_{k,j}(z_{k,t}^{(i)}) = \beta_{k,j} z_{k,t}^{(i)}$ or through nonlinear modeling achieved by applying B-splines. In the latter case, the $k$-th covariate has a B-Spline basis representation $g_{k,j}(z_{k,t}^{(i)}) = \sum_{m=1}^{M_k} \beta_{k,j,m} B_{k,j,m}^l(z_{k,t}^{(i)})$ of order $l\in\mathbb{N}$ with parameter vector $\boldsymbol{\beta}_{k,j} = (\beta_{k,j,1},\dots,\beta_{k,j,M_k})^\top$  (see \citealp{ruppert2003semiparametric} or \citealp{fahrmeir2007regression}). For reasons of identifiability of smooth effects, these functions are centered around zero as proposed in \cite{wood2017generalized}. We collect all regression parameters for state $j$ in a single vector which we denote with $\boldsymbol{\theta}_j$. Estimation of the model parameters is shown in Appendix \ref{app:estimation}.

\section{Results}
\label{sec:Results}

\subsection{Fitting intensities}

\begin{table}[t]
\center
\begin{tabular}{lrc|rc}
\toprule
& \multicolumn{2}{c|}{Model state 0}  & \multicolumn{2}{c}{Model state 1}   \\
\midrule
 & Effect (s.e.) & Relative risk  & Effect (s.e.) & Relative Risk  \\
 \midrule
Intercept & $-2.721$ (0.177) & 0.066 & $-1.751$ (0.146) & 0.174 \\
\midrule
Tuesday & 0.050 (0.012)  & 1.052  &  0.037 (0.012) & 1.037 \\
Wednesday & 0.091 (0.012) & 1.095  &  0.025 (0.012) & 1.025 \\
Thursday & 0.124 (0.012) & 1.132 &  0.005 (0.012) & 1.005\\
Friday & 0.098 (0.012) & 1.103 &  0.024 (0.012) & 1.024\\
Saturday & $-0.139$ (0.013) & 0.870  &  $-0.032$ (0.013) & 0.968 \\
Sunday & 0.126 (0.014) & 1.134 &  $-0.371$ (0.016) & 0.690\\
\midrule
central &  0.159 (0.199) & 1.172 &  $-0.764$ (0.164) & 0.466 \\
south &  0.129 (0.265) & 1.138 &  0.218 (0.218) & 1.244 \\
\midrule
$\texttt{nearby}_{1-j}$ & 1.451 (0.022) & 4.266 & 1.078 (0.023)  & 2.940\\
\bottomrule
\end{tabular}
\caption{Estimates $\widehat{\beta}_{k,j}$ of fixed linear covariate effects in the time to event model with linear predictor as specified in \eqref{eq:linear_predictor}, standard errors in brackets. The relative risk is given as $\exp(\widehat{\beta}_{k,j})$.}
\label{tab:effects}
\end{table}

\begin{figure}
\includegraphics[width = 0.49\textwidth]{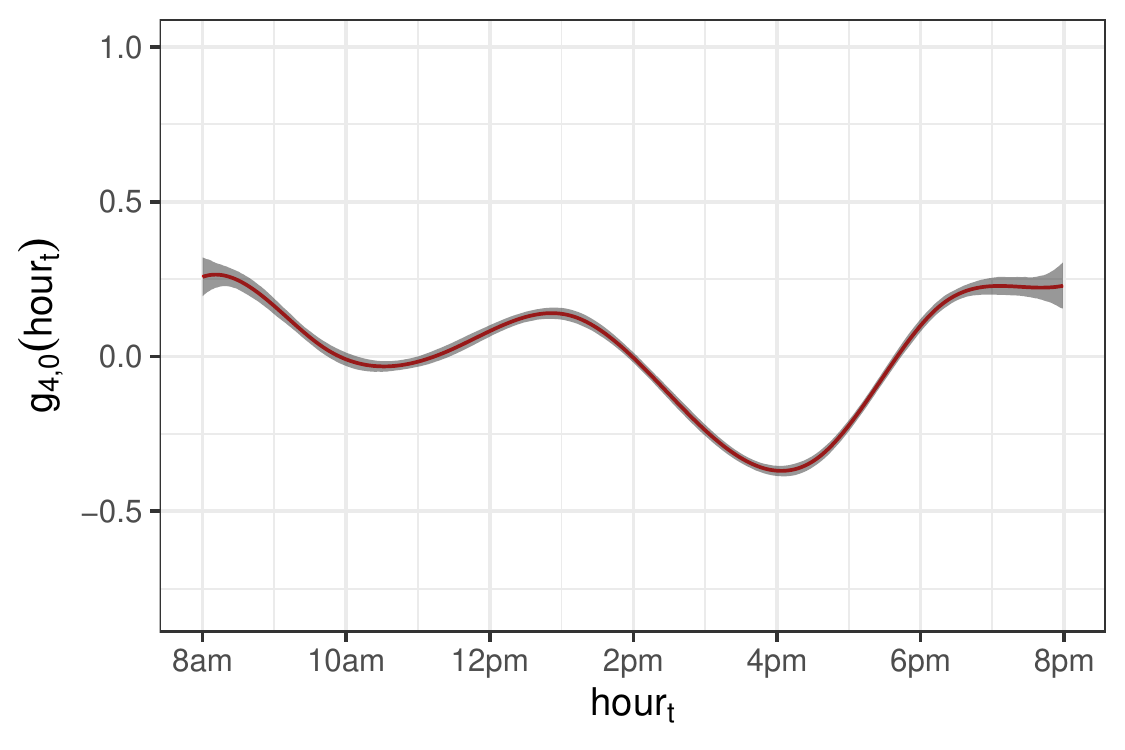}
\includegraphics[width = 0.49\textwidth]{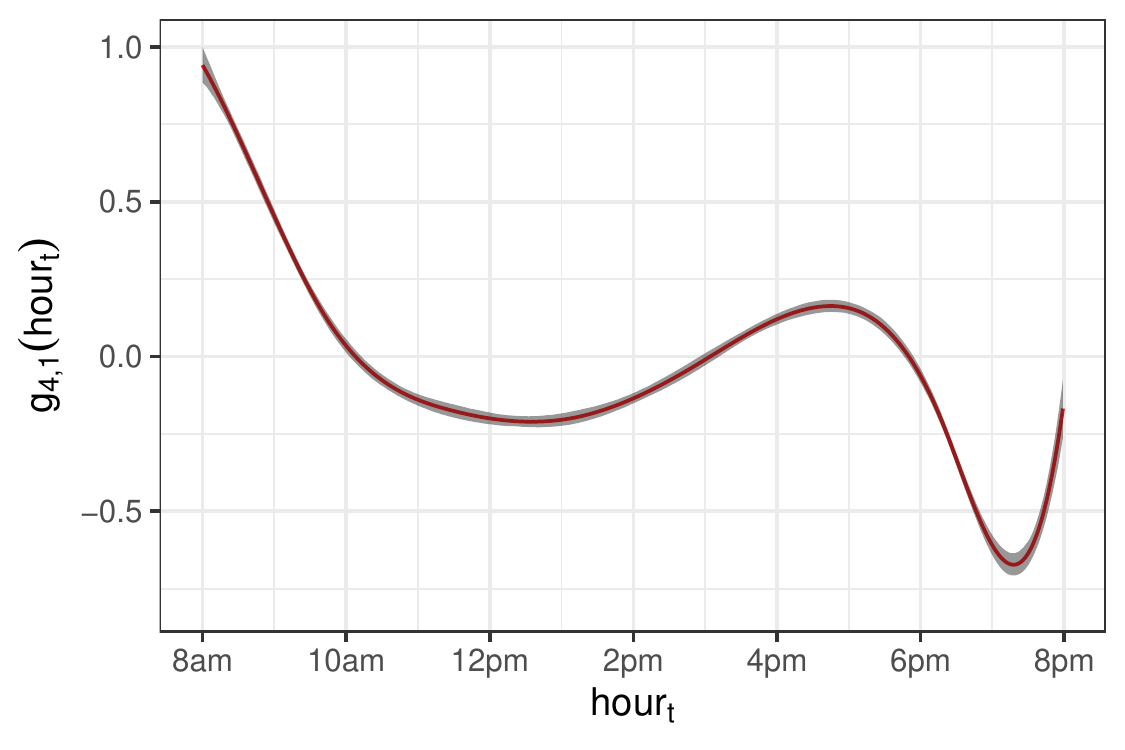}
\caption{Smooth effect of the time of the day in state 0 (left panel) and state 1 (right panel), 95\% confidence bands are shown in grey.}
\label{fig:smooths}
\end{figure}

We first show an exemplary fit of the time to event model from Section \ref{sec:intensities} to the Melbourne parking data which we introduced in Section \ref{sec:data}. The linear predictor for state $j = 0,1$ is specified as
\begin{align}
\eta_{j,t}^{(i)} = \beta_{0,j} + \beta_{1,j} \cdot \texttt{weekday}_t + \beta_{2,j} \cdot \texttt{sideofstreet}^{(i)} + \beta_{3,j} \cdot \texttt{nearby}_{1-j,t}^{(i)} + g_{4,j}(\texttt{hour}_t),
\label{eq:linear_predictor}
\end{align}
where time point $t$ corresponds to the start time of each observed, possibly censored, duration. Temporarily, we here restrict the data to all days of June 2019 between 8 am in the morning and 8 pm in the evening while on the spatial scale we only include parking lots which are located in the eastern section of Lonsdale Street, which is shown as a thicker network segment in Figure \ref{fig:average_duration}. The covariate $\texttt{nearby}_{j,t}^{(i)}$ is defined as
\begin{equation*}
\texttt{nearby}_{j,t}^{(i)} = \sum_{k \neq i} \mathbbm{1} \lbrace X_t^{(i)} = j, d_{\boldsymbol{G}}(\boldsymbol{s}_i, \boldsymbol{s}_k) \leq h \rbrace / \sum_{k \neq i} \mathbbm{1} \lbrace  d_{\boldsymbol{G}}(\boldsymbol{s}_i, \boldsymbol{s}_k) \leq h \rbrace
\end{equation*}
which is the fraction of all parking lots in state $j$ with driving distance less than $h$ from parking lot $i$ at time point $t$. For all our analyses we set $h = 50$ meters. The smooth functions $g_{4,j}(\texttt{hour}_t)$ which model the effect of the time of the day are build with quadratic B-splines with 10 degrees of freedom.

We fit the time to event model which we specified above employing the \texttt{survival} package for the statistical software \textbf{R} \citep{r}. For the estimated shape parameter $\alpha_j$ of the Weibull baseline intensity it holds that $\widehat{\alpha}_1 = 0.55 < \widehat{\alpha}_0 = 0.65 \ll 1$ which mimics strictly decreasing transition intensities. This suits the Kaplan-Meier estimators shown in Figure \ref{fig:KM_lonsdale_east} and legitimates the use of a Weibull baseline hazard instead of an exponential baseline hazard. Estimates of linear covariate effects and their standard errors are shown in Table \ref{tab:effects}. The proportional hazards assumption allows to quantify the relative risk of covariate effects. The effect of the weekly variation is fairly weak except for Sunday, where we see a significant negative effect in the model for the transition from state 1 to state 0. Therefore, the average duration of parking on Sundays is longer which might be caused by more relaxed parking restrictions on Sundays. The effect of the relative location of a parking lot mirrors the conclusion which we drew from Figure \ref{fig:KM_lonsdale_east}, i.e. there is no significant effect in the model for state 0 and the parking duration in between two lanes is expected to be significantly higher than at the curbside. The significant positive effect of the covariate $\texttt{nearby}_{1}$ ($\texttt{nearby}_{0}$) in the model for state 0 (1) means that the duration a parking lot is clear (occupied) decreases with increasing occupancy (availability) of nearby parking lots. Finally, we show in Figure \ref{fig:smooths} the smooth effect of daytime. We see that if a parking lot is cleared in the afternoon the expected duration in state 0 is least. Overall, the effect of the time of the day for state 0 is much weaker when compared to state 1 where the parking duration tendentiously rises during the course of the day, i.e. short-term parking occurs mainly in the morning.

\subsection{Predicting parking lot occupancy}

As we can see, the results of the time to event model are valuable in their own right. However, the main purpose of those was to use them as plug-in estimates for a model in order to predict $\mathbb{P}(X_{t_f}^{(i)} = 0 \mid \mathcal{F})$. Therefore, we next assess the performance of the following three prediction models which were generally treated in Section \ref{sec:semiMarkov}. 
For the first two models we fit a time to event model with Weibull baseline hazard and linear predictor as specified in \eqref{eq:linear_predictor}. The fitted intensities are used in order to estimate the distribution function as derived in \eqref{eq:distribution_function} and the probabilities of interest $\mathbb{P}(X_{t_f}^{(i)} = 0 \mid \mathcal{F})$ are computed according to \eqref{eq:prediction_semiMarkov} where, however, in the first case $d_0$ corresponds to the actually observed value and in the second case $d_0$ will be generally set to zero. Thus, the first model is a semi-Markov model with state space $\mathcal{S}^\star$ of cardinality four while the second model essentially reduces to a semi-Markov model with two states, i.e. state space $\mathcal{S}$. For the third model, we fit the time to event model with auxiliary condition $\alpha_j = 1$ which leads to exponentially distributed duration times. Consequently, the predictions can be made according to the closed-form solution \eqref{eq: P two state Markov}. We refer to this model as the two-state Markov model.
 
In order to evaluate the performance of the different models under a preferably realistic scenario we consider the following setting. We randomly choose a time point $t = 0$ as well as a location $\boldsymbol{s}$ on the street network $\boldsymbol{G}$ (see Figure \ref{fig:average_duration}) where we put a higher sampling weight on areas with more parking lots, see \cite{schneble2020intensity}. Now, we predict the availability of parking lots being clear for all parking lots $i$ which satisfy $d_{\boldsymbol{G}}(\boldsymbol{s}, \boldsymbol{s_i}) \leq 250$ meters. The time point $t = 0$ is chosen to be either between 10 am and 12 pm or between 4 pm and 6 pm of each day in June 2019. The prediction horizon shall be $t_f = 10$ minutes or $t_f = 30$ minutes, respectively. In each case, the transition intensities are determined by making use of data restricted to 30 days in the past from the perspective of $t = 0$.

\begin{figure}
\includegraphics[width = 0.49\textwidth]{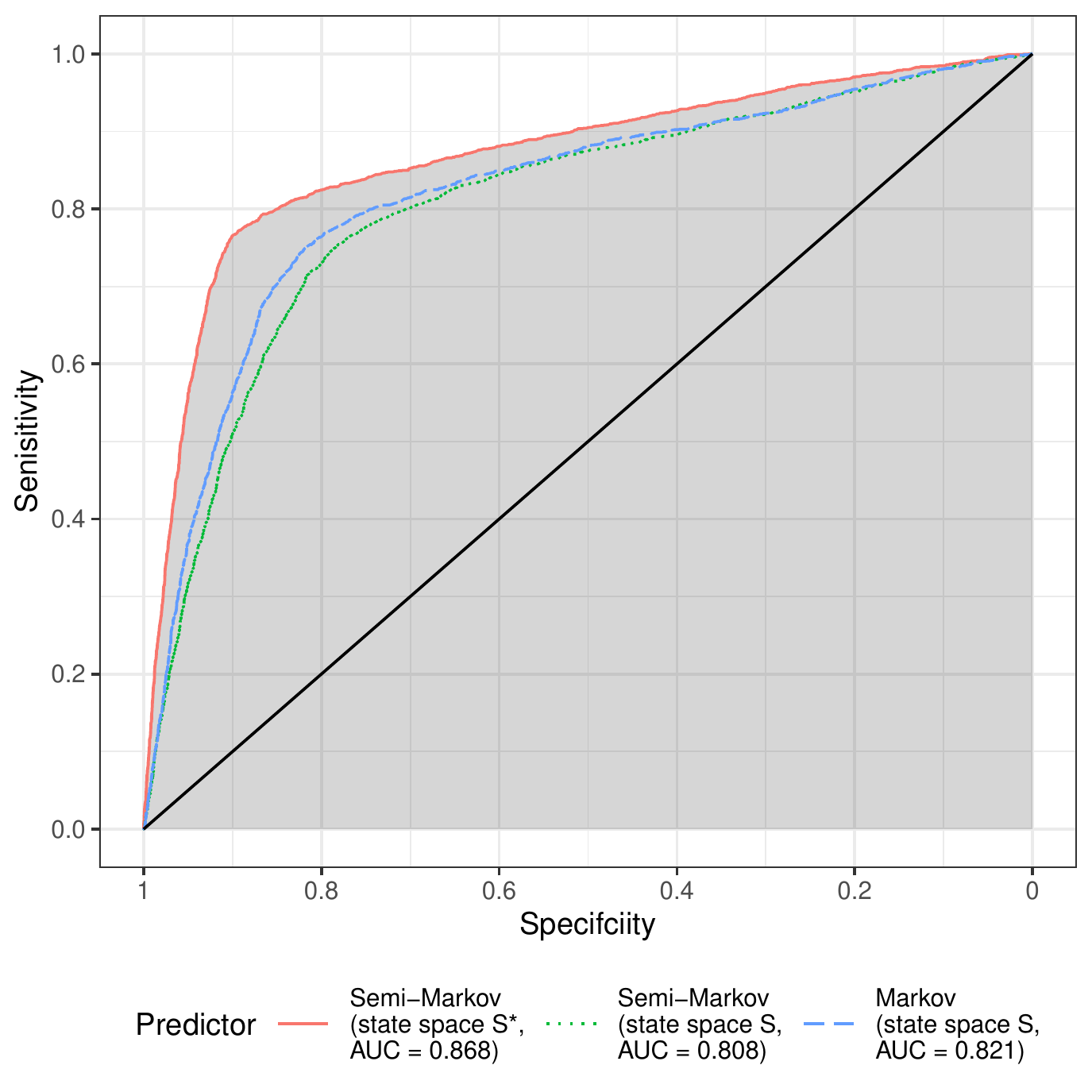}
\includegraphics[width = 0.49\textwidth]{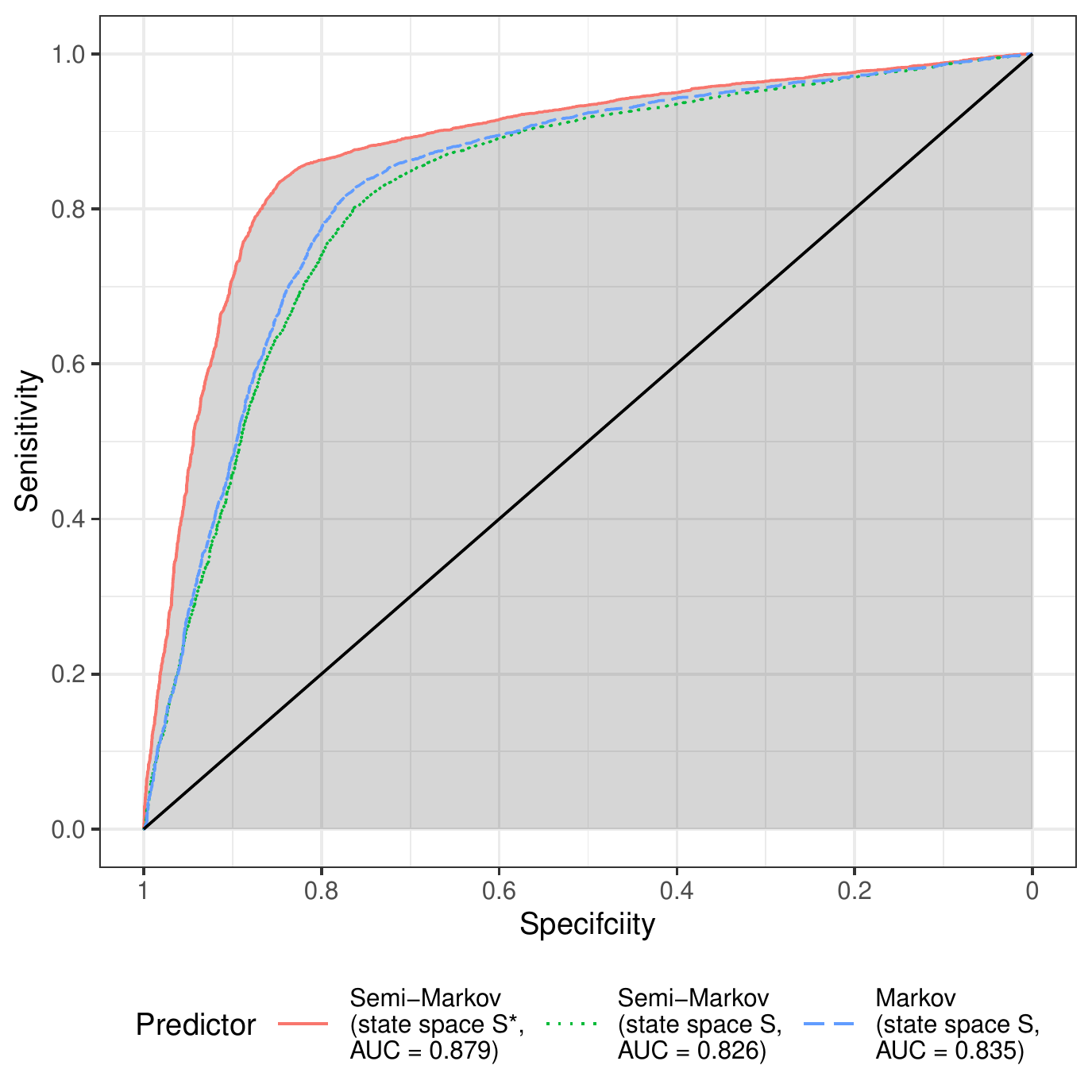} \\
\includegraphics[width = 0.49\textwidth]{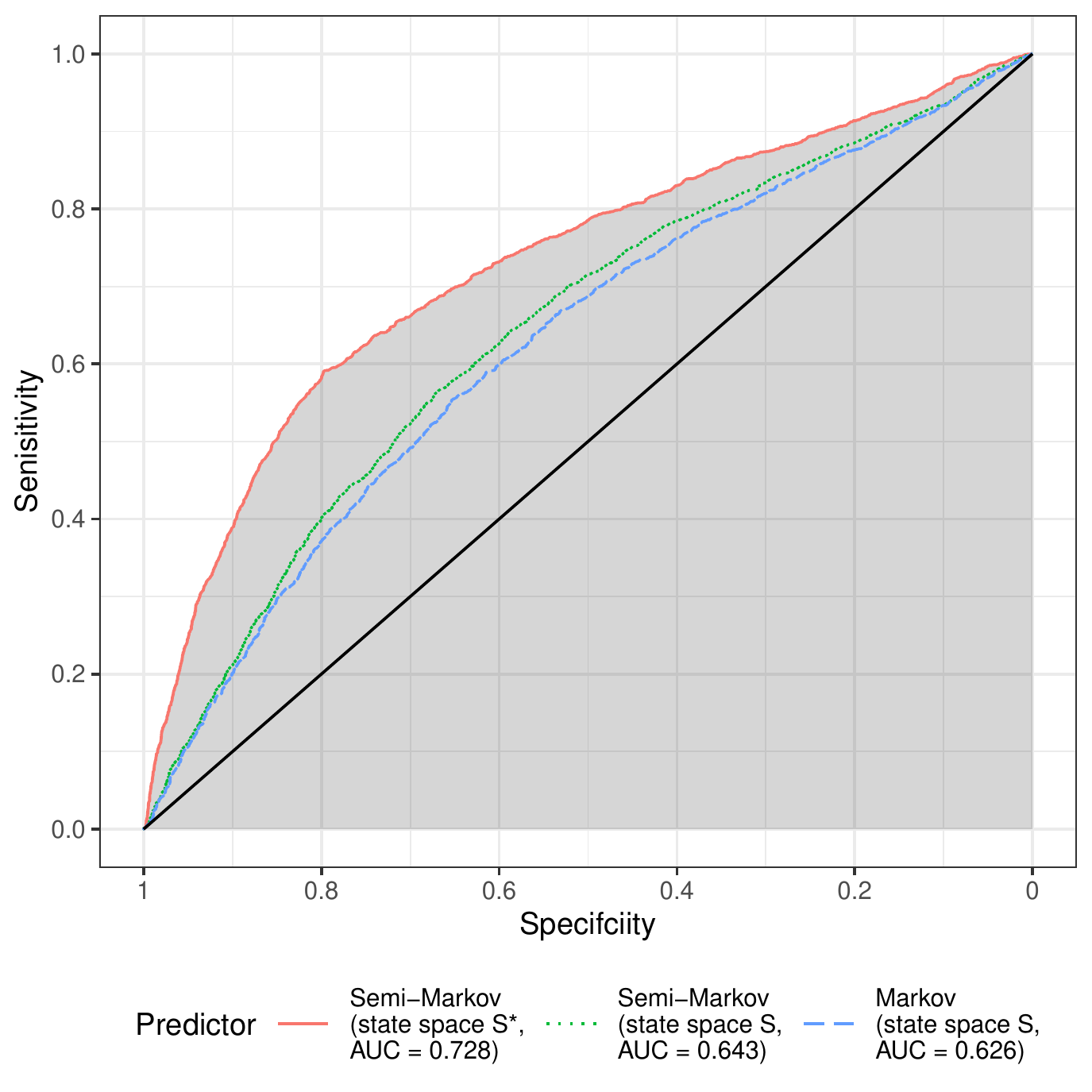}
\includegraphics[width = 0.49\textwidth]{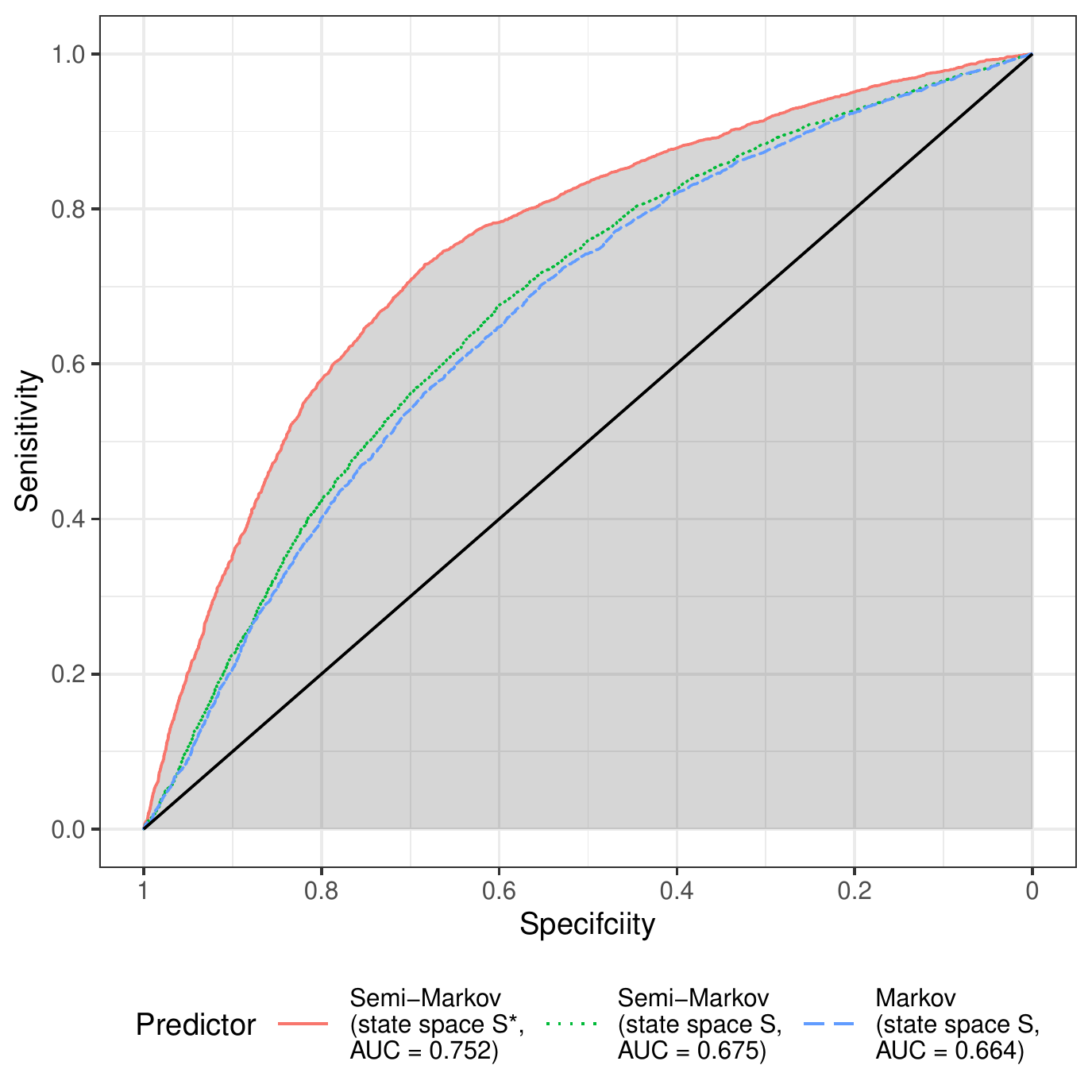}
\caption{ROC curves showing the performance of three predictors when daily predicting free parking lots in June 2019. In the left (right) panels, $t = 0$ ranges between 10 am and 12 pm (4 pm and 6 pm). In the top (bottom) panels, time point $t_f$ is 10 minutes (30 minutes) after $t = 0$.}
\label{fig:ROC}
\end{figure}

We repeat each scenario described above $R = 100$ times and measure the prediction performance by making use of receiver operating characteristic (ROC) curves \citep{robin2011proc}. Thereby, we first specify a set $\lbrace c_p \rbrace$ of $P+2$ thresholds with $ -\infty = c_0 < 0 < c_1 < \dots < c_P < 1 < \infty = c_{P+1}$. Then, we determine for each $c_p$ the binary estimate 
\begin{align*}
\widehat{X}_t^{(i)}(c_p) = \begin{cases}
0, & \mathbb{P}(X_{t_f}^{(i)} = 0 \mid \mathcal{F}) \geq c_p \\
1, & \mathbb{P}(X_{t_f}^{(i)} = 0 \mid \mathcal{F}) < c_p
\end{cases}.
\end{align*} 
Next, we compute the specificity $\text{TNR}(c_p)$ (true negative rate) and the sensitivity $\text{TPR}(c_p)$ (true positive rate) in dependence of the threshold $c_p$ as
\begin{align*}
\text{TNR}(c_p) = \frac{\#\lbrace X_{t_f}^{(i)} = 1 \text{ and } \widehat{X}_{t_f}^{(i)}(c_p) = 1 \rbrace}{\#\lbrace  X_{t_f}^{(i)} = 1 \rbrace}, \quad \text{TPR}(c_p) = \frac{\#\lbrace X_{t_f}^{(i)} = 0 \text{ and } \widehat{X}_{t_f}^{(i)}(c_p) = 0 \rbrace}{\#\lbrace  X_{t_f}^{(i)} = 0 \rbrace}.
\end{align*}
Note that in contrast to the habitual convention, ``0'' refers to the positive class and ``1'' refers to the negative class. Finally, a ROC curve is a function of the sensitivity in dependence of $1 - \text{specificity}$. An index which measures the overall prediction performance of a binary predictor is the area under the (ROC) curve (AUC), which is the integral of the ROC curve. An AUC equal to one corresponds to a perfect predictor where the random predictors AUC, highlighted through a diagonal line in Figure \ref{fig:ROC}, is always equal to 0.5.

In Figure \ref{fig:ROC} we show ROC curves of the predictions related to each of the four possible scenarios described above, i.e. with prediction horizon equal to $t_f =  10$ minutes (upper panels) or $t_f = 30$ minutes (lower panels) and setting the present moment to $t = 0$ in the morning (left panels) or in the late afternoon (right panels). We can generally observe that for a fixed specificity, the semi-Markov predictor with state space $\mathcal{S}^\star$ outperforms the (semi-)Markov predictor with state space $\mathcal{S}$ in terms of sensitivity, where the same holds vice versa. Accordingly, the AUC of the semi-Markov predictor is larger when compared to the AUC of the Markov predictor. This difference is minor if the prediction horizon is very short-term (10 minutes). If the prediction horizon amounts to 30 minutes, the performance of both models worsens distinctly. However, on a relative scale the prediction accuracy of the (semi-)Markov model with two states diminishes stronger when compared to the semi-Markov model with four states. Summarizing, we conclude that the benefit of our methodology mainly comes from the adding of the additional states $0^\star$ and $1^\star$ to the state space $\mathcal{S}$. However, when opposing the two predictors which make use of a two-state stochastic process, it is not apparent that one outperforms the other.

\begin{figure}
\centering
\includegraphics[width = 0.8\textwidth]{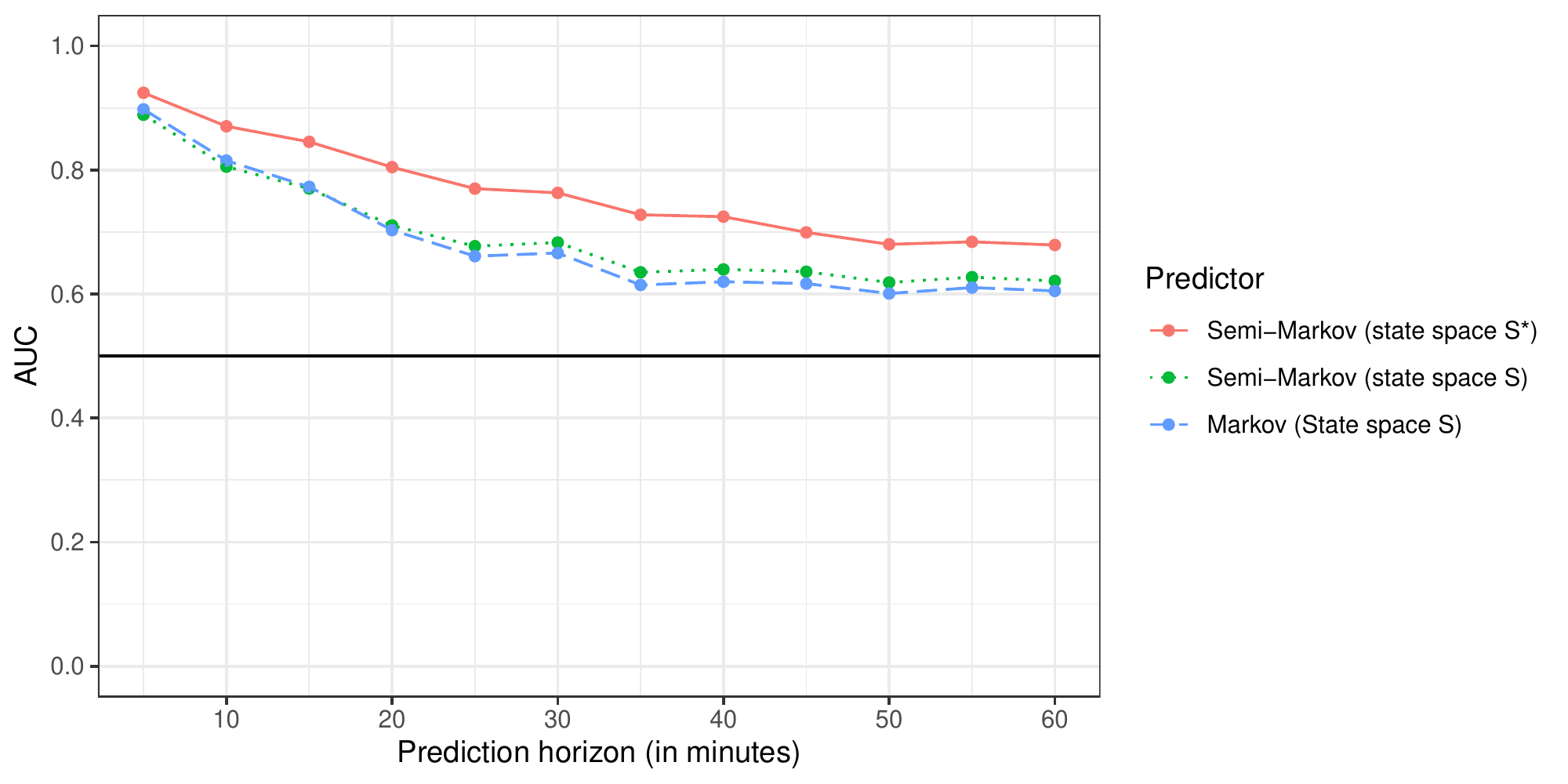}
\caption{AUC of three binary predictors depending on the prediction horizon.}
\label{fig:AUC}
\end{figure}

Finally, we show the AUC for the same predictors as considered in Figure \ref{fig:ROC} when the prediction horizon ranges between five minutes and one hour. Since a large effect of the time of the day on the prediction accuracy can not be observed we draw $t=0$ from the interval between 4 pm and 6 pm. Figure \ref{fig:ROC} confirms what we have seen before, the semi-Markov predictor with the extended state space $\mathcal{S}^\star$ performs superior in terms of AUC when compared to the (semi-)Markov predictor with two states, i.e. when the current duration $d_0$ is not respected in the model. This is most apparent when the prediction horizon is between 20 and 40 minutes. When opposing the semi-Markov predictor with two states to the Markov predictor, we observe only few discrepancy between the two predictors for predictors horizons 20 minutes and less. If the prediction horizon is longer the semi-Markov model can be favored as the AUC is marginally larger. Summing up, the semi-Markov model distinctly outperforms the Markov model in terms of AUC. Again, this is largely due to the involvement of the two additional states.

\section{Discussion}
\label{sec:Discussion}

In this paper we have presented a general framework which can be used to predict the individual short-term probability of on-street parking lots being unoccupied. A time to event model is employed in order to estimate the transition intensities and consequently the distribution of the duration in each state. Besides the usage as plug-in estimates for the (semi-)Markov prediction model, the results of the time to event model already provide valuable insight into the patterns of on-street parking dynamics in the City of Melbourne. On the other hand, the semi-Markov model is solely designed as a prediction model and we have seen that the prediction accuracy, measured in terms of AUC of a ROC curve, is distinctly larger when compared to a model which is restricting duration times to be exponentially distributed. If the response in a binary prediction model is greatly unbalanced, i.e. most of the data belong either to the positive or negative class, \cite{saito2015precision} propose to employ precision-recall curves in order to evaluating a binary predictor. However, we do not see such a strong imbalance in our data such that we consider these results reliable.

The performance of the prediction model is highly dependent on the quality of the data. Usually, not all sensors in a chosen area are active at the same time and not all parking lots are equipped with a sensor. If a sensor does not send the current state of the related parking lot, it is much harder to predict its short-term availability since the initial distribution is not known and particularly not deterministic. Consequently, the initial distribution $\boldsymbol{p}$ of the parking lot occupancy would have to be estimated as well and then $\boldsymbol{P}(t_f) \boldsymbol{p}$ yields a vector of transition probabilities from time $t = 0$ to time $t = t_f$, where $\boldsymbol{P}(t_f)$ can be obtained by applying the inverse Laplace transform to \eqref{eq:solution_complex}.

The time to event model which we employ to estimate the distribution functions of the semi-Markov model is rather unsophisticated. However, this has the advantage that standard software can be used to fit the model. The performance of the prediction model could certainly be further improved by incorporating covariates into the time to event model in a (duration) time varying manner. Moreover, the usage of further external covariates such as weather conditions might increase the prediction performance as well. However, our main intention with this paper was to show that a semi-Markov model clearly outperforms a Markov model in terms of prediction accuracy. Since we use the same estimates for both kinds of prediction models, this finding should be rather independent from the goodness of the chosen time to event model.  

A natural extension of our model is its integration into an individual parking guidance system. Here, a car driver might choose a destination $\boldsymbol{s}$ in the center of an urban area and a navigation system computes the estimated time of arrival $t_f$ until reaching the target $\boldsymbol{s}$. Our model is able to yield predictions of on-street parking availability at time point $t_f$ in the surrounding area of $\boldsymbol{s}$ and could guide the driver into a street section in which the likelihood of finding a clear parking lot is greatest. We consider this computational considerable optimization problem to be beyond the scope of this paper where we focused on the statistical point of view of our modeling approach.

	\bibliographystyle{Chicago}
	
	\bibliography{literature}

\appendix

\section{Parameter estimation in the time to event model}
\label{app:estimation}

When estimating the parameters of the time to event model \eqref{eq:model}, we assume the duration times in states 0 and 1 to be independent, which allows for independent estimation of the model parameters $\boldsymbol{\zeta}_j = (\alpha_j, \boldsymbol{\theta}_j^\top, \gamma_j)^\top$ for $j = 0,1$. Recall that $\gamma_j$ is the parameter which determines the variance of the frailties $v_j^{(i)}$. This seems plausible as the duration of a parking lot being clear should not be related to the duration a parking lot being occupied. This allows for a break of notation in this appendix. We drop state index $j$, parking lot index $i$ is now a subscript instead of a superscript and time index $t$ is understood to be explicitly contained in the linear predictor.

Many procedures for estimating the parameters in a shared frailty model have been discussed. An EM-Algorithm was proposed by \cite{klein1992semiparametric} where the model parameters $\boldsymbol{\zeta}$ and the frailty terms $v_i$ are iteratively estimated. A penalized and likewise iterative estimation algorithm is suggested by  \cite{therneau2003penalized}. However, estimates of $\boldsymbol{\zeta}$ can also be obtained directly by maximizing the marginal log-likelihood of the model. Following the arguments in \cite{kalbfleisch2011statistical}, for the conditional likelihood related to the $i$-th parking lot it holds that
\begin{align}
L_i(\alpha, \boldsymbol{\theta} \mid v_i) = \prod_{k=1}^{n_i} \left[ \alpha d_{ik}^{\alpha} v_i \exp\left(\eta_{ik}\right) \right]^{\delta_{ik}}  \exp \left( - \exp\left(\eta_{ik} \right)  v_i \left(d_{ik}\right)^{\alpha} \right).
\label{eq:conditional_likelihood}
\end{align}
Here, index $k$ refers to the $k$-th observed duration time of a parking lot and $n_i$ is the number of duration times observed for parking lot $i$. Moreover, $\delta_{ik} = 1$ if the corresponding event is observed and $\delta_{ik} = 0$ if the duration time $d_{ik}$ is right censored. Note that we censor all duration times which are longer than 60 minutes which has the desired consequence that $d_{ik}$ and $\delta_{ik}$ are independent. The marginal likelihood of the model can be obtained by integrating the frailty terms out of \eqref{eq:conditional_likelihood} for every $i$ 
\begin{align}
L_{\text{marg}}(\boldsymbol{\zeta}) &= \prod_{i=1}^N \int_0^\infty L_{i}(\alpha, \boldsymbol{\theta} \mid v_i)  h(v_j^{(i)}; \gamma) \dif v_i
\label{eq:marginal_likelihood}
\intertext{with}
h(v; \gamma) &= \frac{v^{\frac{1}{\gamma}-1}}{\gamma^\frac{1}{\gamma}\Gamma(\frac{1}{\gamma})}\exp\left(-\frac{v}{\gamma}\right)\mathbbm{1}_{(0,\infty)}(v)
\notag
\end{align}
denoting the density of the $\text{Gamma}(\frac{1}{v},\frac{1}{v})$ distribution. In this special case the integral in \eqref{eq:marginal_likelihood} is analytically tractable  and the marginal log-likelihood of the model results to
\begin{align*}
\ell_{\text{marg}}(\boldsymbol{\zeta}) &= \sum_{i=1}^N \left[ \Delta_i \log \gamma + \log \frac{\Gamma\left(\frac{1}{\gamma} + \Delta_i \right)}{\Gamma\left(\frac{1}{\gamma}\right)}    \vphantom{\sum_{k=1}^{n_i}}   \right. -  \left(\frac{1}{\gamma} + \Delta_i\right) \log \left(  1  + \gamma \sum_{k=1}^{n_i} \left(d_{ik}\right)^{\alpha}  \right) \\
 &+ \left. \sum_{k=1}^{n_i}  \delta_{ik}\left( \eta_{ik} + \log \alpha + (\alpha-1) \log d_{ik}  \right) \right],
\end{align*} 
where $\Delta_i = \sum_{k=1}^{n_i} \delta_{ik}$  and the maximum likelihood estimate of the fixed parameters is given by $\boldsymbol{\widehat{\zeta}} = \text{argmax } \ell_{\text{marg}}(\boldsymbol{\zeta})$ \citep{duchateau2007frailty}. Finally, we need predicted values for the random effects $u_i$ which are based on the posterior mean of the frailties $v_i = \log u_i$ given the observations $\boldsymbol{d}_i = (d_{i1},\dots,d_{in_i})^\top$. Using Bayes' theorem, it can be shown that the posterior density $h(v \mid \boldsymbol{d}_{i}; \widehat{\boldsymbol{\zeta}})$ of $v_i$ is equal to the density of a gamma distribution with shape parameter $a_i = \frac{1}{\widehat{\gamma}} + \Delta_i$ and scale parameter $b_i = \frac{1}{\widehat{\gamma}} + \sum_{k=1}^{n_i} \widehat{\alpha} \left(d_{ik}\right)^{\widehat{\alpha}-1} \exp \left( \widehat{\eta}_{ik} \right)$  and therefore, $\widehat{v_i} = \mathbb{E}(v_i \mid \boldsymbol{d}_i; \widehat{\boldsymbol{\zeta}}) = \frac{a_i}{b_i}$ \citep{nielsen1992counting}.

\section{Numerical inversion of the Laplace transform}
\label{app:invlap}

We here provide details about the algorithm which we employ to numerically compute the inverse Laplace transform of a Laplace transformed function $\widetilde{f} = \mathcal{L}\lbrace f \rbrace : C \rightarrow \mathbb{C}$, where $C \in \mathbb{C}$ is the region of convergence. The complete derivation of the underlying theoretical results can be found in \cite{valsa1998approximate}. Recall that the inverse Laplace transform is defined as the following Bromwich integral
\begin{align}
f(t) = \mathcal{L}^{-1} \lbrace \widetilde{f}(u)\rbrace(t) = \frac{1}{2\pi i} \int_{\gamma-i\infty}^{\gamma+i\infty} \widetilde{f}(u) e^{ut} \dif u,
\label{eq:app1}
\end{align}
where $u = x+iy$ and $\gamma > \min\lbrace \mathfrak{Re}(u) \mid u \in C \rbrace$. The basic idea of the algorithm is to approximate the complex exponential in \eqref{eq:app1} as
\begin{align}
e^{ut} &\approx \frac{e^a}{2\sinh(a-ut)} = \frac{e^{ut}}{1-e^{-2(a-ut)}}
\notag
\intertext{such that}
f(t) \approx f(t;a) &= f(t) + \sum_{n=1}^\infty e^{-2na}f((2n+1)t) \notag \\
&= \frac{e^a}{t} \sum_{n=0}^\infty (-1)^n2^{-[\mathbbm{1}_{\lbrace 0 \rbrace}(n)]} \mathfrak{Re}\left[   \widetilde{f}\left(  \frac{a+in\pi}{t} \right) \right]
\label{eq:app2}
\end{align}
where $a > 0$ is a tuning parameter. It is recommended to choose $a = 6$ \citep{valsa1998approximate}. Naturally, the sum in \eqref{eq:app2} needs to be truncated at some positive integer $n_t$. In order to increase the speed of convergence, \cite{valsa1998approximate} propose to keep $n_t$ rather low, but adding an Euler approximation of the subsequent $n_e$ summands. Consequently, a finite numerical approximation for \eqref{eq:app1} is
\begin{align*}
f(t) \approx f(t;a) &\approx \frac{e^{a}}{t}\sum_{n=0}^{n_t} (-1)^n2^{-[\mathbbm{1}_{\lbrace 0 \rbrace}(n)]} \mathfrak{Re}\left[   \widetilde{f}\left(  \frac{a+in\pi}{t} \right) \right] \\
&+ \frac{e^a2^{-n_e}}{t}\sum_{n=n_t+1}^{n_t+n_e}(-1)^{n} \left[ \sum_{k=n-n_t}^{n_e} \binom{n_e}{k} \right] \mathfrak{Re}\left[   \widetilde{f}\left(  \frac{a+in\pi}{t} \right)  \right].
\end{align*}

\end{document}